\documentclass{aastex}

\usepackage{graphicx}
\usepackage{amsmath}
\usepackage{amsfonts}
\usepackage{amssymb}
\usepackage{epsfig}
\usepackage{float}
\usepackage{makeidx}
\usepackage{mathrsfs}
\usepackage{subfig}

\newcommand{\begeq}{\begin{equation}}
\newcommand{\fineq}{\end{equation}}
\newcommand{\begeqarray}{\begin{eqnarray}}
\newcommand{\fineqarray}{\end{eqnarray}}
\newcommand{\xmin}{x_{\rm min}}
\newcommand{\xmax}{x_{\rm max}}
\newcommand{\green}{f_{_{\rm G}}}

\def\green{f_{_{\rm G}}}
\def\Green{F_{_{\rm G}}}
\def\sigmaT{\sigma_{\rm T}}
\def\xmax{x_{\rm max}}
\def\xmin{x_{\rm min}}

\def\sgreen{f_{_{\rm G}}^{\rm S}}

\newcommand{\gapprox}{\lower.4ex\hbox{$\;\buildrel
>\over{\scriptstyle\sim}\;$}}
\newcommand{\lapprox}{\lower.4ex\hbox{$\;\buildrel
<\over{\scriptstyle\sim}\;$}}

\slugcomment{submitted to ApJ}

\shorttitle{X-Ray Time Lags in a Comptonizing Corona}

\shortauthors{Kroon, Becker}

\begin{document}

\title{AN INTEGRATED MODEL FOR THE PRODUCTION OF X-RAY TIME LAGS
AND QUIESCENT SPECTRA FROM HOMOGENEOUS AND INHOMOGENEOUS BLACK HOLE
ACCRETION CORONAE}

\author{John J. Kroon}

\affil{School of Physics, Astronomy, and Computational Sciences,
George Mason University, Fairfax, VA 22030-4444, USA; jkroon@gmu.edu}

\author{Peter A. Becker}

\affil{School of Physics, Astronomy, and Computational Sciences,
George Mason University, Fairfax, VA 22030-4444, USA; pbecker@gmu.edu}

\begin{abstract}
Many accreting black holes manifest time lags during outbursts, in which
the hard Fourier component typically lags behind the soft component.
Despite decades of observations of this phenomenon, the underlying
physical explanation for the time lags has remained elusive, although
there are suggestions that Compton reverberation plays an important role.
However, the lack of analytical solutions has hindered the interpretation
of the available data. In this paper, we investigate the generation of X-ray time lags in
Compton scattering coronae using a new mathematical approach based on
analysis of the Fourier-transformed transport equation. By solving this equation,
we obtain the Fourier transform of the radiation Green's function, which allows
us to calculate the exact dependence of the time lags on the Fourier frequency,
for both homogeneous and inhomogeneous coronal clouds.
We use the new formalism to explore a variety of injection scenarios, including both
monochromatic and broadband (bremsstrahlung) seed photon injection.
We show that our model can successfully reproduce both the
observed time lags and the time-averaged (quiescent) X-ray spectra for Cyg~X-1 and GX~339-04,
using a single set of coronal parameters for each source. The time
lags are the result of impulsive bremsstrahlung injection occurring
near the outer edge of the corona, while the time-averaged spectra are the result
of continual distributed injection of soft photons throughout the cloud.

\end{abstract}



\section{INTRODUCTION}

Many accretion-powered X-ray sources display rapid variability, coupled
with a time-averaged spectrum consisting of a power law terminating in
an exponential cutoff at high energies. The ubiquitous nature of the
observations suggests a common mechanism for the spectral formation
process, regardless of the type of central object (e.g. black hole,
neutron star, AGN, etc.). Over the past few decades, the interpretation of
the spectral data using steady-state models
has demonstrated that the power-law component is most likely due to the
thermal Comptonization of soft seed photons in a hot ($\sim 10^8\,$K) coronal
cloud (Sunyaev \& Titarchuk 1980). While the spectral models yield
estimates for the coronal temperature and optical depth, they do not provide
much detailed information about the geometry and morphology of the plasma.
On the other hand, observations of variability, characterized by time lags and
power spectral densities (PSDs), can supplement the spectral analysis, yielding
crucial additional information about the structure of the inner region in the
accretion flow, where the most rapid variability is generated.

In particular, the study of X-ray time lags, in
which the hard photons associated with a given Fourier component arrive
at the detector before or after the soft photons, provides a unique
glimpse into the nature of the high-frequency variability in the inner
region. Fourier time lags offer an ideal tool for studying rapid
variability because, unlike short-timescale spectral snapshots, which
become noisy due to the shortage of photons in small time bins, the
Fourier technique utilizes all of the data in the entire observational
time window, which could extend over hundreds or thousands of seconds.
Hence the resulting time lag information usually has much higher
significance than can be achieved using conventional spectral
analysis.

\subsection{Fourier Time Lags}

The Fourier method for computing time lags from observational data
streams in two energy channels was pioneered by van der Klis et al.
(1987), who proposed a novel mathematical technique for extracting time
lags by creating a suitable combination of the hard and soft Fourier
transforms for a given value of the circular Fourier frequency, $\omega$. The
method utilizes the Complex Cross-Spectrum, denoted by $C(\omega)$,
defined by
\begin{equation}
C(\omega) \equiv S^*(\omega) \, H(\omega)
\label{eq1}
\ ,
\end{equation}
where $S$ and $H$ are the Fourier transforms of the soft and hard
channel time series, $s(t)$ and $h(t)$, respectively, and $S^*$ denotes
the complex conjugate. The Fourier transforms are calculated using
\begin{equation}
S(\omega) = \int_{-\infty}^{\infty}e^{i\omega t} s(t)dt
\ ,
\label{eq2}
\end{equation}
\noindent and likewise for the hard channel,
\begin{equation}
H(\omega) = \int_{-\infty}^{\infty}e^{i\omega t} h(t)dt
\label{eq3}
\ .
\end{equation}
The phase lag between the two data streams is computed by taking the
argument of $C(\omega)$, which is the argument angle in the complex plane,
and the associated time lag, $\delta t$, is obtained by dividing the
phase lag by the Fourier frequency. Hence we have the relations
\begin{equation}
\delta t = \frac{{\rm arg}(C)}{2\pi \nu_f}
= \frac{{\rm arg}(S^{*}H)}{2\pi \nu_f}
\ ,
\label{eq4}
\end{equation}
where the Fourier frequency, $\nu_f$, is related to the circular frequency
$\omega$ via
\begin{equation}
\nu_f = \frac{\omega}{2 \pi}
\ .
\label{eq4b}
\end{equation}

As a simple demonstration of the time lag concept, it is instructive to consider
the case where the hard and soft channels, $h(t)$ and $s(t)$, are shifted in time
by a precise interval $\Delta t$, so that the two signals are related to each via
\begin{equation}
h(t) = s(t-\Delta t)
\label{eq5}
\ ,
\end{equation}
where $\Delta t > 0$ would indicate a hard time lag. Next we take the Fourier
transform of the hard channel time series to obtain
\begin{equation}
H(\omega) = \int_{-\infty}^{\infty}e^{i\omega t}h(t)dt
= \int_{-\infty}^{\infty}e^{i\omega t}s(t-\Delta t)dt
\label{eq6}
\ .
\end{equation}
Introducing a new time variable, $t'=t-\Delta t$ with $dt'=dt$, allows
us to transform the integral in Equation~(\ref{eq6}) to obtain
\begin{equation}
H(\omega) = \int_{-\infty}^{\infty}
e^{i\omega(t'+\Delta t)}s(t')dt'
= e^{i\omega \Delta t}S(\omega)
\ .
\label{eq7}
\end{equation}
It follows from Equation~(\ref{eq1}) that the resulting complex
cross-spectrum is given by
\begin{equation}
C(\omega) = S^*(\omega)e^{i\omega \Delta t}S(\omega)
= e^{i\omega \Delta t} |S(\omega)|^2
\label{eq8}
\ ,
\end{equation}
and hence the resulting time lag is (cf. Equation~(\ref{eq4}))
\begin{equation}
\delta t = \dfrac{\omega \Delta t}{\omega}
= \Delta t
\ .
\label{eq9}
\end{equation}
This simple calculation confirms that the time lag computed using the
Fourier method gives the correct answer when a perfect delay is
introduced between the two channels, as expected. It is also important
to note that time lags are only produced during a transient. We can see
this by setting the hard and soft signals equal to the constants $h_0$
and $s_0$, respectively, so that $h(t)=h_0$ and $s(t)=s_0$. In this
case, the resulting Fourier transforms $H$ and $S$ have the same phase,
and consequently there is no phase lag or time lag. Hence observations
of time lags necessarily imply the presence of variability in the
observed signal.

\subsection{X-Ray Time Lag Phenomenology}

The fundamental physical mechanism underlying the X-ray time lag
phenomenon has been debated for decades, but it is generally accepted
that the time lags reflect the time-dependent scattering of a population of
seed photons that are impulsively injected into an extended corona
of hot electrons (e.g., van der Klis et al. 1987; Miyamoto et al. 1988). This initial
population of photons gain energy as they Comptonize in the cloud, and
the hard time lags are a natural consequence of the extra time that the
hard photons spend in the cloud gaining energy via electron scattering
before escaping. In contrast with the time lags, the time-averaged (quiescent) spectra
are thought to be created as a result of the Compton scattering of
{\it continually} injected seed photons. The time-dependent upscattering
of soft input photons is discussed in detail by Payne (1980) and
Sunyaev \& Titarchuk (1980), who present fundamental formulas
for the resulting X-ray spectrum. Since that time, many detailed
models have been proposed, most of which focus on a single
aspect of radiative transfer, usually by making assumptions about the
physical conditions in the disk/corona system regarding the
electron temperature, the input photon spectrum, and the size
and optical depth of the scattering corona.

The Fourier time lags observed from accreting black-hole sources
generally decrease with increasing Fourier frequency, $\nu_f$. In
the case of Cyg~X-1, for example, the time lags decrease from
$\sim 0.1-10^{-3}$~sec as $\nu_f$ increases from
$\sim 0.1\,$Hz - $10^2\,$Hz . Early attempts to
interpret this data using simple Compton scattering models resulted in
very large, hot scattering clouds, which required very efficient heating
at large distances ($\sim 10^{5-6}~GM/c^2$) from the central mass
(Poutanen \& Fabian 1999, Hua et al. 1999, hereafter HKC). Furthermore,
the observed dependence of the time lags on the Fourier frequency was difficult to
explain using a homogeneous Compton scattering model. For example, van der Klis et
al. (1987) and Miyamoto et al. (1988) found that a homogeneous corona
combined with monochromatic soft photon injection resulted in time
lags that are {\it independent} of the Fourier frequency, $\nu_f$, in contradiction
to the observations. This led Miyamoto et al. (1988) to conclude, somewhat
prematurely, that thermal Comptonization could not be
producing the lags. However, in the next decade, HKC and Nowak et al. (1999)
developed more robust Compton simulations that successfully reproduced the observed
time lags, although the large coronal radii $\sim 10^{4.5-5.5}~GM/c^2$
continued to raise concerns regarding energy conservation and heating.

HKC computed the time lags and the time-averaged spectra for a variety of
electron number density profiles, based on the injection of
low-temperature blackbody seed photons at the center of the coronal
cloud. They employed a two-region structure, comprising a central
homogeneous zone, connected to a homogeneous or inhomogeneous outer
region that extends out to several light-seconds from the central mass.
In the inhomogeneous case, the electron number density, $n_e(r)$, in the outer
region varied as $n_e(r) \propto r^{-1}$ or $n_e(r) \propto r^{-3/2}$. In the HKC
model, the injection spectrum and the injection location were both held
constant, and a zero-flux boundary condition was adopted at the center
of the cloud. HKC found that only the model with $n_e(r) \propto r^{-1}$
in the outer region was able to successfully reproduce the observed
dependence of the time lags on the Fourier frequency. On the other
hand, in the homogeneous case, HKC confirmed the Miyamoto et al.
(1988) result that the time lags are independent of the Fourier frequency,
in contradiction to the observational data. This result was also verified later
by Kroon \& Becker (2014, hereafter KB) for the case of monochromatic
photon injection into a homogenous corona.

\subsection{Dependence on Injection Model}

Despite the progress made by HKC and other authors, no successful
first-principles theoretical model for the production of the observed X-ray time lags
has yet emerged. In the absence of such a model, one is completely dependent
on Monte Carlo simulations, which are somewhat inconvenient since the resulting
time lags are not analytically connected with the parameters describing
the scattering cloud. Monte Carlo simulations are also noisy at high Fourier frequency,
which is the main region of interest in many applications, although this can be dealt
with by adding more test particles. Compared with an analytical calculation,
the utilization of Monte Carlo simulations makes it more challenging to explore
different injection scenarios, such as the variation of the injection location
and the seed photon spectrum.

The situation changed recently with the work of KB, who presented a
detailed analytical solution to the problem of time-dependent
thermal Comptonization in spherical, homogeneous scattering clouds. By
obtaining the fundamental {\it photon Green's function} solution to the
problem, they were able to explore a wide variety of injection
scenarios, leading to a better understanding of the relationship between the
observed time lags and the underlying physical parameters.
KB verified the Miyamoto result, namely that monochromatic
injection in a homogeneous cloud produces time lags that are independent
of Fourier period. The magnitude of this (constant) lag depends primarily on the
radius of the cloud, $R$, its optical thickness, $\tau_*$, and the
electron temperature, $T_e$. Following HKC, they employed a zero-net flux boundary
condition at the center of the corona (essentially a mirror condition),
so that injection could occur at any radius inside the cloud. The photon
transport at the outer edge of the cloud was treated using a free-streaming
boundary condition in order to properly account for photon escape. KB demonstrated
that the injection radius and the shape of the injected photon spectrum play
a crucial role in determining the dependence of the resulting time lags on
the Fourier frequency. In particular, they established for the first time that the
reprocessing of a broadband injection spectrum (e.g., thermal bremsstrahlung)
can successfully reproduce most of the time lag data for Cyg~X-1 and other
sources.

In the study presented here, we expand on the work of KB to obtain the
radiation Green's function for {\it inhomogeneous} scattering clouds. We also present
a more detailed derivation of the {\it homogeneous} Green's function discussed
by KB. The analytical solutions for the Fourier transform of the time-dependent
Green's function in the homogeneous and inhomogeneous cases are then
used to treat localized bremsstrahlung injection via integral convolution,
as an alternative to the essentially monochromatic injection scenario
studied by HKC. In addition to modeling the transient time lags as a result
of {\it impulsive} soft photon injection, we also compute the time-independent
X-ray spectrum radiated form the surface of the cloud as a result of {\it continual}
soft photon injection. We show that acceptable fits to both the time-lag data
and the X-ray spectral data can be obtained using a single set of cloud
parameters (temperature, density, cloud radius) via application of our integrated
model.

The remainder of the paper is organized as follows. In
Section~2 we introduce the time-dependent and steady-state transport
equations in spherical geometry, and we map out the general solution
methods to be applied in the subsequent sections. In Section~3 we obtain the
solution for the Fourier transform of the time-dependent photon Green's
function and also the solution for the time-averaged
Green's function in a homogeneous corona. In Section~4, we repeat the same
steps for the case of an inhomogeneous corona with electron number density
profile $n_e(r) \propto 1/r$. We discuss the reprocessing of thermal bremsstrahlung
radiation in Section~5, and we apply the integrated model to Cyg~X-1 and GX~339-04
in Section~6. Our main conclusions are reviewed and further discussed in Section~7.

\section{Fundamental Equations}

Our focus here is on understanding how time-dependent Compton scattering
affects a population of seed photons as they propagate through a spherical corona
of hot electrons overlying a geometrically thin, standard accretion disk. This problem
was first explored using an exact mathematical approach by KB, who studied the
radiative transfer occurring in a homogeneous corona. We provide further details
of that work here, and we also extend the model to treat inhomogeneous spherical
scattering clouds.

\subsection{Time-Dependent Transport Equation}

The time-dependent transport equation describing the diffusion and
Comptonization of an instantaneous flash of $N_0$ monochromatic seed
photons injected with energy $\epsilon_0$ at radius $r_0$ and at time
$t_0$ as they propagate through a spherical scattering corona is given
by (e.g., Becker 2003),
\begin{multline}
\dfrac{\partial \green}{\partial t} = \dfrac{1}{r^2}
\dfrac{\partial }{\partial r}\left[\kappa(r) \, r^2 \dfrac{
\partial \green}{\partial r}\right] + \dfrac{n_e(r) \sigmaT c}
{m_e c^2} \dfrac{1}{\epsilon^2}\dfrac{\partial }{\partial
\epsilon}\left[\epsilon^4\left(\green + k T_e\dfrac{\partial
\green}{\partial \epsilon}\right)\right] \\
+ \dfrac{N_0 \delta(t-t_0)\delta(r-r_0)\delta(\epsilon-\epsilon_0)}
{4\pi r_0^2\epsilon_0^2}
\ ,
\label{eq10}
\end{multline}
where $m_e$, $n_e$, $T_e$, $k$, $\sigmaT$, $c$, and $\kappa$
denote the electron mass, the electron number density, the electron
temperature, Boltzmann's constant, the Thomson cross section, the
speed of light, and the spatial diffusion coefficient, respectively, and
$\green(\epsilon,r,t)$ is the radiation Green's function,
describing the distribution of photons inside the cloud. The first term
on the right-hand side of Equation~(\ref{eq10}) represents the spatial
diffusion of photons through the corona, and the second term describes
the redistribution in energy due to Compton scattering. The Green's
function is related to the photon number density, $n_r$, via
\begin{equation}
n_r(r,t) = \int_0^\infty \epsilon^2 \, \green(\epsilon,r,t) \, d\epsilon
\label{eq11}
\ ,
\end{equation}
and the spatial diffusion coefficient $\kappa(r)$ is related to the
electron number density $n_e(r)$ and the scattering mean free
path $\ell(r)$ via
\begin{equation}
\kappa(r) = \dfrac{c}{3n_e(r)\sigmaT} = {c \, \ell(r) \over 3}
\label{eq12}
\ . 
\end{equation}
Klein-Nishina corrections are important when the incident photon energy
in the electron's rest frame approaches $\sim 500\,$keV. In our model, the electrons
are essentially non-relativistic, with temperature $T_e \sim 4-7 \times 10^8\,$K, and therefore
the $0.1-10\,$keV photons of interest here will not be boosted into the Klein-Nishina
energy range in the typical electron's rest frame. We will therefore treat the
electron scattering process using the Thomson cross section throughout this study.
However, we revisit this issue is Section~7.1 where we compare our results with
previous studies that utilized the full Klein-Nishina cross section to treat the electron
scattering.

\subsection{Density Variation}

In many cases of interest, the electron number density $n_e(r)$ has a power-law
dependence on the radius $r$, which can be written as
\begin{equation}
n_e(r)=n_*\left(\dfrac{r}{R}\right)^{-\alpha}
\label{eq13}
\ ,
\end{equation}
where $R$ is the outer radius of the cloud, $\alpha$ is a constant, and
$n_* \equiv n_e(R)$ is the number density at the outer edge of the
cloud.
The two cases we focus on here are
\begin{equation}
\alpha =
\begin{cases}
0, & {\rm homogeneous} \ , \\
1, & {\rm inhomogeneous} \ .
\end{cases}
\label{eq14}
\end{equation}
The homogeneous case was treated by Miyamoto (1988) and the inhomogeneous
case by HKC. By combining Equations~(\ref{eq12}) and (\ref{eq13}), we
can rewrite the electron number density and the spatial diffusion
coefficient as
\begin{equation}
n_e(r) = {1 \over \sigmaT \ell_*} \left(r \over R\right)^{-\alpha} \ ,
\ \ \ \ \ \
\kappa(r) = {c \, \ell_* \over 3 }\left({r \over R}\right)^{\alpha}
\ ,
\label{eq15}
\end{equation}
where
\begin{equation}
\ell_*\equiv \ell(R) = \frac{1}{n_e(R) \sigmaT}
\label{eq15.1}
\end{equation}
denotes the scattering mean free path
at the outer edge of the corona. Substituting Equations~(\ref{eq15})
into Equation~(\ref{eq10}) yields
\begin{multline}
\dfrac{\partial \green}{\partial t} = \dfrac{c \ell_*}{3 r^2}
\dfrac{\partial }{\partial r}\left[\left(\dfrac{r}{R}
\right)^{\alpha} r^2 \dfrac{\partial \green}{\partial r}
\right] + \dfrac{1}{\ell_* m_e c}\left(\dfrac{r}{R}
\right)^{-\alpha} \dfrac{1}{\epsilon^2}\dfrac{\partial}
{\partial \epsilon} \left[\epsilon^4\left(\green + kT_e
\dfrac{\partial \green}{\partial \epsilon}\right)\right] \\
+ \dfrac{N_0\delta(t-t_0)\delta(r-r_0)
\delta(\epsilon-\epsilon_0)}{4\pi r_0^2\epsilon_0^2}
\ .
\label{eq16}
\end{multline}

The electron temperature $T_e$ is determined by a balance between
gravitational heating and Compton cooling, and one typically finds that
$T_e$ does not vary significantly in the region where most of the X-rays
are produced (You et al. 2012; Schnittman et al. 2013). We therefore assume
that the cloud is isothermal with $T_e=$~constant. In this case, it is
convenient to rewrite the transport equation in terms of the
dimensionless energy
\begin{equation}
x \equiv \dfrac{\epsilon}{kT_e}
\ .
\label{eq17}
\end{equation}
We also introduce the dimensionless radius $z$, time $p$, and temperature
$\Theta$, defined, respectively, by
\begin{equation}
z \equiv \dfrac{r}{R} \ , \ \ \ \ \
p \equiv \dfrac{c \, t}{\ell_*} \ , \ \ \ \ \
\Theta \equiv \dfrac{k T_e}{m_e c^2}
\ .
\label{eq18}
\end{equation}
The various functions involved in the derivation can be written in terms of
either the dimensional energy and radius, $(\epsilon,r)$, or the corresponding
dimensionless variables $(x,z)$, and therefore we will use these two
notations interchangeably throughout the remainder of the paper.
Incorporating Equations~(\ref{eq17}) and (\ref{eq18}) into
the transport equation~(\ref{eq16}) yields, after some
algebra,
\begin{equation}
\dfrac{\partial \green}{\partial p} =
{1 \over 3 \eta^2 z^2} \dfrac{\partial}
{\partial z}\left(z^{2+\alpha} \dfrac{\partial \green}
{\partial z}\right)
+ \dfrac{\Theta}{z^\alpha x^2}\dfrac{\partial}{\partial x}
\left[x^4\left(\green+\dfrac{\partial \green}{\partial x}
\right)\right]
+ \dfrac{N_0\delta(x-x_0)\delta(p-p_0)\delta(z-z_0)}
{4\pi z_0^2 R^3 x_0^2 \Theta^3 (m_e c^2)^3}
\ ,
\label{eq19}
\end{equation}
where we have introduced the dimensionless ``scattering parameter,''
\begin{equation}
\eta \equiv \dfrac{R}{\ell_*} = n_e(R) \sigmaT R
\ .
\label{eq20}
\end{equation}
Equation~(\ref{eq19}) is the fundamental partial differential
equation that we will use to treat time-dependent scattering in a
homogeneous spherical corona with $\alpha=0$ in Section~3, and
time-dependent scattering in an inhomogeneous spherical corona
with $\alpha=1$ in Section~4.

\subsection{Optical Depth}

The scattering optical depth $\tau$ measured from the inner edge
of the coronal cloud at radius $r=r_{\rm in}$ out to some arbitrary
local radius $r$ is computed using
\begin{equation}
\tau(r) = \int_{r_{\rm in}}^r n_e(r') \sigmaT dr'
= \int_{r_{\rm in}}^r {dr' \over \ell(r')}
\ ,
\label{eq21}
\end{equation}
where the variation of the mean-free path is given by (see
Equations~(\ref{eq12}) and (\ref{eq13}))
\begin{equation}
\ell(r) = \ell_* \left({r\over R}\right)^\alpha
\ .
\label{eq22}
\end{equation}
Combining relations, and transforming the variable of integration
from $r$ to $z=r/R$, we obtain
\begin{equation}
\tau(z) = \eta \int_{z_{\rm in}}^z \, {dz' \over z'^\alpha}
\label{eq23}
\ ,
\end{equation}
where
\begin{equation}
z_{\rm in} \equiv {r_{\rm in} \over R}
\label{eq24}
\end{equation}
denotes the dimensionless inner radius of the cloud.

There are three cases of interest here,
\begin{equation}
\tau(z) =
\begin{cases}
\eta \, (z^{1-\alpha} - z_{\rm in}^{1-\alpha})/(1-\alpha) \ ,
& \alpha \ne 1 \ , \\
\eta \, (z - z_{\rm in}) \ , & \alpha = 0 \ , \\
\eta \, \ln (z/z_{\rm in}) \ , & \alpha = 1 \ .
\end{cases}
\label{eq25}
\end{equation}
The overall optical thickness of the scattering cloud, denoted by $\tau_*$, as
measured from the inner radius $r=r_{\rm in}$ ($z=z_{\rm in}$) to the outer radius
$r=R$ ($z=1$), is therefore given by
\begin{equation}
\tau_* =
\begin{cases}
\eta \, (1 - z_{\rm in}^{1-\alpha})/(1-\alpha) \ ,
& \alpha \ne 1 \ , \\
\eta \, (1 - z_{\rm in}) \ , & \alpha = 0 \ , \\
\eta \, \ln (1/z_{\rm in}) \ , & \alpha = 1 \ .
\end{cases}
\label{eq26}
\end{equation}

\subsection{Steady-State Transport Equation}

The time-averaged (quiescent) X-ray spectra produced in accretion flows
around black holes are generally interpreted as the result of the
thermal Comptonization of soft seed photons continually injected into a
hot electron corona from a cool underlying disk (see e.g. Sunyaev \&
Titarchuk 1980 for a review). In our interpretation, the associated
X-ray time lags are the result of the time-dependent Comptonization of
seed photons impulsively injected during a brief transient. Our goal in
this paper is to develop an integrated model that accounts for the
formation of both the time-averaged spectrum and the time lags using a
single set of cloud parameters (temperature, density, radius). In our
calculation of the time-averaged spectrum, we assume that $\dot N_0$ seed
photons with energy $\epsilon_0$ are injected per unit time into the hot
corona between the inner cloud radius $r_{\rm in}$ and the outer cloud
radius $r=R$ with a rate that is proportional to the local electron
number density $n_e(r)$. The radial variation of the number density
depends on whether the cloud is homogeneous, with $n_e=$constant,
or inhomogeneous, with $n_e(r) \propto r^{-1}$.

In this scenario, the fundamental time-independent transport equation
can be written as
\begin{equation}
\dfrac{\partial \sgreen}{\partial t} = 0 = \dfrac{1}{r^2}
\dfrac{\partial }{\partial r}\left[\kappa(r) r^2 
\dfrac{\partial \sgreen}{\partial r}\right] +
\dfrac{n_e(r) \sigmaT c}{m_e c^2} \dfrac{1}{\epsilon^2}
\dfrac{\partial} {\partial \epsilon}\left[\epsilon^4
\left(\sgreen + kT_e\dfrac{\partial \sgreen}{\partial
\epsilon}\right)\right] + \dfrac{\dot{N}_0 \,
\delta(\epsilon-\epsilon_0)n_e(r)} {\epsilon_0^2 N_e}
\label{eq27}
\ ,
\end{equation}
where $\sgreen(\epsilon,r)$ denotes the steady-state (quiescent) photon
Green's function, and
\begin{equation}
N_e = \int_{r_{\rm in}}^{R} 4\pi r^2 n_e(r) \, dr
\label{eq28}
\end{equation}
represents the total number of electrons in the region $r_{\rm in} \le r
\le R$. Substituting for $n_e(r)$ and $\kappa(r)$ in Equation~(\ref{eq27})
using Equations~(\ref{eq15}) yields
\begin{equation}
0 = \dfrac{c \ell_*}{3 r^2}
\dfrac{\partial }{\partial r}\left[\left(\dfrac{r}{R}
\right)^{\alpha} r^2 \dfrac{\partial \sgreen}{\partial r}
\right] + \dfrac{1}{\ell_* m_e c}\left(\dfrac{r}{R}
\right)^{-\alpha} \dfrac{1}{\epsilon^2}\dfrac{\partial}
{\partial \epsilon} \left[\epsilon^4\left(\sgreen + kT_e
\dfrac{\partial \sgreen}{\partial \epsilon}\right)\right] \\
+ \dfrac{\dot{N}_0 \, \delta(\epsilon-\epsilon_0) (r/R)^{-\alpha}}
{\sigmaT \ell_* \epsilon_0^2 N_e}
\ .
\label{eq29}
\end{equation}
This expression can be rewritten in terms of the dimensionless parameters
$x$, $z$, $\Theta$, and $\eta$ to obtain
\begin{equation}
0 = \dfrac{1}{3\eta^2 z^{2-\alpha}}\dfrac{\partial }
{\partial z}\left(z^{2+\alpha} \dfrac{\partial \sgreen}{\partial z}\right)
+\dfrac{\Theta}{x^2}\dfrac{\partial}{\partial x} \left[x^4\left(\sgreen
+\dfrac{\partial \sgreen}{\partial x}\right)\right]
+\dfrac{\dot{N}_0 \,\delta(x-x_0)(3-\alpha)}
{4\pi R^2 \eta c \, \Theta^3 (m_e c)^3 x_0^2 (1-z_{\rm in}^{3-\alpha})}
\ ,
\label{eq30}
\end{equation}
where we have also substituted for $N_e$ using
\begin{equation}
N_e = {4 \pi R^3 \over \sigmaT \ell_*} \
{1-z_{\rm in}^{3-\alpha} \over 3-\alpha}
\ ,
\label{eq31}
\end{equation}
which follows from Equations~(\ref{eq15}) and (\ref{eq28}). We assume
here that $\alpha=0$ or $\alpha=1$.

The derivative $\partial\sgreen/\partial x$ exhibits a step-function
discontinuity at the injection energy, $x= x_0$, due to the appearance
of the function $\delta(x-x_0)$ in Equation~(\ref{eq30}).
By integrating Equation~(\ref{eq30}) with respect to $x$ over a small
region surrounding the injection energy, we conclude that the derivative
jump is given by
\begin{equation}
\lim_{\delta \rightarrow 0}\left[\dfrac{d\sgreen}
{dx}\right]\Bigg|_{x_0-\delta}^{x_0+\delta}
= -\dfrac{\dot N_0 (3-\alpha)}{4\pi R^2 \eta c \, \Theta^4 (m_e c)^3 x_0^4
(1-z_{\rm in}^{3-\alpha})}
\ .
\label{eq32}
\end{equation}
We will utilize Equations~(\ref{eq30}) and (\ref{eq32}) in Sections~3 and 4
when we compute the time-averaged X-ray spectra produced via electron
scattering in homogeneous and inhomogeneous scattering coronae, respectively.

\subsection{Fourier Transformation}

In principle, all of the detailed spectral variability due to
time-dependent Comptonization in the scattering corona can be computed
by solving the fundamental transport equation (\ref{eq19}) for a given
initial photon energy/space distribution (Becker 2003). However,
complete information about the variability of the spectrum is not
required, or even desired, if the goal it to compare the theoretically
predicted time lags $\delta t$ with the observational data. Computation
of the predicted time lags using Equation~(\ref{eq4}) requires as input
the Fourier transforms of the soft and hard data streams. It is
therefore convenient to analyze the time-dependent transport
Equation~(\ref{eq19}) {\it directly in the Fourier domain}, rather than
in the time domain. Hence one of our goals is to derive the exact
solution for the Fourier transform, $\Green$, of the time-dependent
radiation Green's function, $\green$. We define the Fourier transform
pair, $(\green,\Green)$, using
\begin{equation}
\Green(x,z,\tilde{\omega}) \equiv
\int_{-\infty}^\infty e^{i \tilde{\omega} p}
\green(x,z,p) \, dp
\ ,
\label{eq33}
\end{equation}
\begin{equation}
\green(x,z,p) \equiv \dfrac{1}{2 \pi} \int_{-\infty}^\infty
e^{-i \tilde{\omega} p} \Green(x,z,\tilde{\omega}) \, d\tilde{\omega}
\label{eq34}
\ ,
\end{equation}
where the dimensionless Fourier frequency is defined by
\begin{equation}
\tilde{\omega}=\omega \left(\dfrac{\ell_*}{c}\right)=\omega t_*
\ .
\label{eq35}
\end{equation}
Here, $t_*=\ell_*/c$ is the ``scattering time,'' which equals the mean-free
time at the outer edge of the corona, at radius $r=R$.

We can obtain an ordinary differential equation satisfied by the Fourier transform,
$\Green$, by operating on Equation~(\ref{eq19}) with $\int_{-\infty}^\infty
e^{i\tilde{\omega} p}dp$, to obtain
\begin{multline}
-i \tilde{\omega}z^{\alpha} \Green = \dfrac{1}{3\eta^2 z^{2-\alpha}}
\dfrac{\partial}{\partial z} \left(z^{2+\alpha} \dfrac{\partial
\Green}{\partial z}\right) + \dfrac{\Theta}{x^2} \dfrac{\partial}{\partial x}
\left[x^4\left(\Green + \dfrac{\partial \Green}{\partial x}\right)\right] \\
+ \dfrac{N_0\delta(x-x_0)\delta(z-z_0)e^{i \tilde{\omega} p_0}}{4\pi
x_0^2 z_0^2 z^{-\alpha} \Theta^3 (m_e c^2)^3 R^3}
\ ,
\label{eq36}
\end{multline}
where $i^2=-1$. Further progress can be made by noting that
Equation~(\ref{eq36}) is separable in the energy and spatial
coordinates $(x,z)$. The technical details depend on the value of
$\alpha$, which determines the spatial variation of the electron number
density $n_e(r)$. We therefore treat the homogeneous and inhomogeneous
cases separately in Sections~3 and 4, respectively.

Due to the function $\delta(x-x_0)$ appearing in the source term in
Equation~(\ref{eq36}), the energy derivative $\partial\Green/\partial x$
displays a jump at the injection energy $x= x_0$, with a magnitude determined
by integrating Equation~(\ref{eq36}) with respect to $x$ in a small region around
the injection energy. The result obtained is
\begin{equation}
\lim_{\delta \rightarrow 0}\left[\dfrac{d\Green}{dx}
\right]\Bigg|_{x_0-\delta}^{x_0+\delta} = -\dfrac{N_0\,\delta(z-z_0)
e^{i \tilde{\omega} p_0}}{4\pi x_0^4 \, z_0^2 z^{-\alpha} \, \Theta^4
(m_e c^2)^3 R^3}
\ .
\label{eq37}
\end{equation}
This expression will be used later in the computation of the expansion
coefficients for the Fourier transform of the radiation Green's function resulting
from time-dependent Comptonization in Sections~3.2 and 4.2.

\subsection{Boundary Conditions}

In order to obtain solutions for $\sgreen(\epsilon,r)$ and $\Green(\epsilon,r,\tilde\omega)$,
we must impose suitable spatial boundary conditions at the inner edge of
the cloud, $r=r_{\rm in}$, and at the outer edge, $r=R$, which correspond
to the dimensionless radii $z=z_{\rm in}$ and $z=1$, respectively. The boundary
conditions we discuss below are stated in terms of the fundamental
time-dependent photon Green's function, $\green(\epsilon,r,t)$, but they also apply
to the time-averaged spectrum $\sgreen(\epsilon,r)$. Furthermore, we can show via
Fourier transformation that the same boundary conditions also apply
to the Fourier transform $\Green(\epsilon,r,\tilde\omega)$. Note  that we can
write the time-averaged X-ray spectrum $\sgreen$ and the Fourier transform
$\Green$ as functions of either the dimensional energy and radius, $(\epsilon,r)$,
or in terms of the dimensionless variables $(x,z)$, and therefore we will use the
appropriate set of variables depending on the context.

In the Monte Carlo simulations performed by HKC, the time lags result
from the reprocessing of blackbody seed photons impulsively injected at
the center of the Comptonizing corona. In order to avoid unphysical
sources or sinks of radiation at the center of the cloud, $r=0$, they
employed a zero-flux ``mirror'' inner boundary condition, which can
be expressed as
\begin{equation}
\lim_{r \rightarrow 0} \, -4\pi r^2 \kappa(r) \dfrac{\partial
\green(\epsilon,r,t)}{\partial r} = 0
\label{eq38}
\ .
\end{equation}
This condition simply reflects the fact that no photons are created or
destroyed at the center of the cloud after the initial flash.
Following HKC, we will employ the mirror boundary condition at the
center of the corona ($r=0$) in our calculations involving a homogeneous
cloud.

The scattering corona has a finite extent, and therefore we must impose
a free-streaming boundary condition at the outer surface ($r=R$). Hence the
distribution function $\green$ must satisfy the outer boundary condition
\begin{equation} 
-\kappa(r)\dfrac{\partial \green(\epsilon,r,t)}{\partial r}\Bigg|_{r=R}
= c\,\green(\epsilon,r,t)\Bigg|_{r=R}
\label{eq39}
\ ,
\end{equation}
which implies that the diffusion flux at the surface is equivalent to
the outward propagation of radiation at the speed of light.

When the electron distribution is inhomogeneous ($n_e(r) \propto r^{-1}$),
the mirror condition cannot be applied at the center of the cloud due to
the divergence of the electron number density $n_e(r)$ as $r \to 0$. In
this case, we must truncate the scattering corona at a non-zero inner
radius, $r=r_{\rm in}$, where we impose a free-streaming boundary condition.
Physically, the inner edge of the cloud may correspond to the edge of a
centrifugal funnel, or the cusp of a thermal condensation feature
(Meyer \& Meyer-Hofmeister 2007). The inner free-streaming boundary
condition can be written as
\begin{equation} 
-\kappa(r)\dfrac{\partial \green(\epsilon,r,t)}{\partial r}
\Bigg|_{r=r_{\rm in}}
= -c\,\green(\epsilon,r,t)\Bigg|_{r=r_{\rm in}}
\label{eq40}
\ ,
\end{equation}
which is only applied in the inhomogeneous case. All of the boundary conditions
considered here are satisfied by the fundamental time-dependent photon Green's
function $\green(\epsilon,r,t)$, and also by the time-averaged spectrum
$\sgreen(\epsilon,r)$, and the Fourier transform $\Green(\epsilon,r,\tilde\omega)$.
We will apply these results in Sections~3 and 4 where we consider homogeneous
and inhomogeneous cloud configurations, respectively.

\section{Homogeneous Model}

The simplest electron number density distribution of interest here is
$n_e=$constant ($\alpha=0$), which was first studied by Miyamoto et al.
(1988). In this case we apply the mirror inner boundary condition at the
center of the cloud, and hence we set $z_{\rm in}=0$. We consider the
homogeneous case in detail in this section, and obtain the exact
solutions for the Fourier transform of the time-dependent photon Green's
function, $\Green(\epsilon,r,\tilde\omega)$, and also for the associated
time-averaged radiation spectrum, $\sgreen(\epsilon,r)$. These results were
originally presented by KB in an abbreviated form. Note that KB utilized
the scattering optical depth $\tau$ measured from the center of the cloud
as the fundamental spatial variable, whereas we use the dimensionless
radius $z$. However, the two quantities are simply related via
Equations~(\ref{eq25}) and (\ref{eq26}), which yield, for $\alpha=0$
and $z_{\rm in}=0$,
\begin{equation}
\tau(z) = \eta \, z \ , \ \ \ \ \
\tau_* = \eta
\ ,
\label{eq41}
\end{equation}
where $\tau_*$ is the optical thickness measured from the center of the
cloud to the outer edge at $z=1$.

\subsection{Quiescent Spectrum for $\alpha=0$}

In the homogeneous case ($\alpha=0$), the time-independent transport
equation~(\ref{eq30}) representing the thermal Comptonization of seed
photons continually injected throughout the scattering corona can be
simplified by substituting the separation functions
\begin{equation}
f_{\lambda} = K(\lambda,x) \ Y(\lambda,z)
\label{eq42}
\ ,
\end{equation}
which yields, for $x \neq x_0$,
\begin{equation}
\dfrac{-1}{Y \eta^2 z^2}\dfrac{d}
{dz}\left(z^2 {dY \over dz}\right) = {3 \, \Theta \over K \, x^2}{d \over dx}
\left[x^4\left(K + {dK \over dx}\right)\right] = \lambda
\label{eq43}
\ ,
\end{equation}
where $\lambda$ is the separation constant. The corresponding ordinary
differential equations satisfied by the spatial and energy functions $Y$
and $K$ are, respectively,
\begin{equation}
{1 \over z^2}\dfrac{d}
{dz}\left(z^{2} {dY \over dz}\right) + \lambda \, \eta^2 Y = 0
\label{eq44}
\ ,
\end{equation}
\begin{equation}
{1 \over x^2}\dfrac{d}{dx} \left[x^4\left(K + {dK \over dx}\right)
\right]-\dfrac{\lambda}{3\Theta} \, K = 0
\label{eq45}
\ ,
\end{equation}
which has been considered previously by such authors as Payne (1980), 
Shapiro, Lightman, and Eardley (1976), Sunyaev \& Titarchuk (1980), etc.

The fundamental solution for the energy function $K$ is given by (see
Becker 2003)
\begin{equation}
K(\lambda,x) = (x x_0)^{-2} e^{-(x+x_0)/2} M_{2,\sigma}(x_{\rm min})
W_{2,\sigma}(x_{\rm max})
\label{eq46}
\ ,
\end{equation}
where $M_{2,\sigma}$ and $W_{2,\sigma}$ are Whittaker functions,
\begin{equation}
x_{\rm max} \equiv \max(x,x_0) \ , \ \ \ \ \
x_{\rm min} \equiv \min(x,x_0)
\label{eq47}
\ ,
\end{equation}
and
\begin{equation}
\sigma \equiv \sqrt{\dfrac{9}{4}+\dfrac{\lambda}{3\Theta}}
\ .
\label{eq48}
\end{equation}
The specific form in Equation~(\ref{eq46}) represents the solution
satisfying appropriate boundary conditions at high and low energies, and
it is also continuous at the injection energy, $x=x_0$, as required.

In the homogeneous configuration under consideration here, the spatial
function $Y$ must satisfy the inner ``mirror'' boundary condition at the
origin (cf. Equation~(\ref{eq38})), which can be written in terms of
$z$ as
\begin{equation}
\lim_{z \rightarrow 0} \ z^2 \, {dY(\lambda,z) \over dz} = 0
\label{eq49}
\ .
\end{equation}
The fundamental solution for $Y$ satisfying this condition is given by
\begin{equation} 
Y(\lambda,z) =
{\sin(\eta z\sqrt{\lambda}) \over \eta z}
\label{eq50}
\ .
\end{equation}
By virtue of Equation~(\ref{eq39}), the  spatial function $Y$ must
also satisfy the outer free-streaming boundary condition, written in terms
of the $z$ coordinate as
\begin{equation}
\lim_{z \rightarrow 1} \left[{1 \over 3\eta} {dY(\lambda,z) \over dz}
+ Y(\lambda,z)\right] = 0
\label{eq51}
\ .
\end{equation}
Substituting the form for $Y$ given by Equation~(\ref{eq50}) into
Equation~(\ref{eq51}) yields a transcendental equation for the
eigenvalues $\lambda_n$ that can be solved using a numerical
root-finding procedure. The resulting eigenvalues $\lambda_n$ are all
real and positive, and the corresponding values of $\sigma$ are computed
by setting $\lambda=\lambda_n$ in Equation~(\ref{eq48}). The
associated eigenfunctions, $Y_n$ and $K_n$, are defined by
\begin{equation}
Y_n(z) \equiv Y(\lambda_n,z) \ , \ \ \ 
K_n(x) \equiv K(\lambda_n,x)
\label{eq52}
\ .
\end{equation}

According to the Sturm-Liouville theorem, the eigenfunctions $Y_n$ form
an orthogonal basis with respect to the weight function $z^2$, so that
(see Appendix~A)
\begin{equation}
\int_0^1 z^2 \, Y_n(z) \, Y_m(z) \, dz = 0 \ , \ \ \ \ \ n \ne m
\label{eq53}
\ .
\end{equation}
The related quadratic normalization integrals, $\mathscr{I}_n$, are
defined by
\begin{equation}
\mathscr{I}_n \equiv \eta^3 \int_0^1 z^2 Y_n^2(z) dz
= {\eta \over 2} - {\sin(2 \eta \sqrt{\lambda_n}) \over 4 \sqrt{\lambda_n}}
\ ,
\label{eq54}
\end{equation}
where the final result follows from Equation~(\ref{eq50}).

Based on the orthogonality of the $Y_n$ functions, we can express
the time-averaged photon Green's function using the expansion
\begin{equation}
\sgreen(x,x_0,z) = \sum_{n=0}^{\infty} b_n \, K_n(x) \, Y_n(z)
\label{eq55}
\ ,
\end{equation}
where the expansion coefficients $b_n$ are computed using the
derivative jump condition in Equation~(\ref{eq32}). In the case
of interest here, we set $\alpha=0$ and $z_{\rm in}=0$ to obtain
\begin{equation}
\lim_{\delta \rightarrow 0}\left[\dfrac{d\sgreen}
{dx}\right]\Bigg|_{x_0-\delta}^{x_0+\delta}
= - \dfrac{3 \dot N_0}{4\pi R^2 \eta c \, \Theta^4 (m_e c)^3 x_0^4}
\ .
\label{eq56}
\end{equation}
Substituting the series expansion for the steady-state Green's function
(Equation~(\ref{eq55})) into Equation~(\ref{eq56}) yields
\begin{equation}
\lim_{\delta \rightarrow 0}\sum_{n=0}^\infty
b_n Y_n(z)[K_n'(x_0+\delta)-K_n'(x_0-\delta)]
= -\dfrac{3 \dot N_0}{4\pi R^2 \eta c \, \Theta^4 (m_e c)^3 x_0^4}
\ .
\label{eq57}
\end{equation}

We can make further progress by eliminating $K$ using Equation~(\ref{eq46})
to obtain, after some algebra,
\begin{equation}
\sum_{n=0}^\infty
b_n \, Y_n(z) \, \mathscr{W}_{2,\sigma}(x_0)
= -\dfrac{3 \dot N_0 e^{x_0}}{4\pi R^2 \eta c \, \Theta^4 (m_e c)^3}
\label{eq58}
\ ,
\end{equation}
where we have defined the Wronskian of the Whittaker functions using
\begin{equation}
\mathscr{W}_{2,\sigma}(x_0) \equiv M_{2,\sigma}(x_0)W'_{2,\sigma}(x_0)
- W_{2,\sigma}(x_0)M'_{2,\sigma}(x_0) 
\ .
\label{eq59}
\end{equation}
The Wronskian can be evaluated analytically to obtain (Abramowitz
\& Stegun 1970)
\begin{equation}
\mathscr{W}_{2,\sigma}(x_0) = - {\Gamma(1+2\sigma) \over \Gamma(\sigma-3/2)}
\label{eq60}
\ .
\end{equation}

Combining Equations~(\ref{eq58}) and (\ref{eq60}), we obtain
\begin{equation}
\sum_{n=0}^{\infty} b_n Y_n(z) \, {\Gamma(1+2\sigma)
\over \Gamma(\sigma-3/2)} = \dfrac{3 \dot N_0 e^{x_0}}{4\pi R^2 \eta c
\, \Theta^4 (m_e c^2)^3}
\ .
\label{eq61}
\end{equation}
Next we exploit the orthogonality of the $Y_n$ functions with respect
to the weight function $z^2$ by applying the operator $\int_0^1 \eta^3
z^2 Y_m(z) dz$ to both sides of Equation~(\ref{eq61}). According to
Equation~(\ref{eq53}), all of the terms on the left-hand side vanish
except the term with $m=n$. The result obtained for the expansion
coefficient $b_n$ is therefore
\begin{equation}
b_n = \dfrac{3 \dot N_0 e^{x_0}\Gamma(\sigma-3/2)\mathscr{P}_n}
{4\pi R^2 \eta c \, \Theta^4 (m_e c^2)^3 \Gamma(1+2\sigma)\mathscr{I}_n}
\label{eq62}
\ ,
\end{equation}
where the integrals $\mathscr{I}_n$ are computed using Equation~(\ref{eq54}) and
the integrals $\mathscr{P}_n$ are defined by
\begin{equation}
\mathscr{P}_n \equiv \int_0^1 \eta^3 z^2 Y_n(z) dz
= {3 \eta \sin(\eta \sqrt{\lambda_n}) \over \lambda_n}
\label{eq63}
\ ,
\end{equation}
and the final result follows from application of Equation~(\ref{eq51}).

Combining Equations~(\ref{eq55}) and (\ref{eq62}) yields the exact
analytical solution for the time-independent photon Green's function evaluated
at dimensionless energy $x$ and dimensionless radius $z$ resulting
from the continual injection of seed photons throughout the cloud. We obtain
\begin{equation}
\sgreen(x,x_0,z) = {9 \dot N_0 e^{x_0} \over 4 \pi R^2 c \, \Theta^4 (m_e c^2)^3}
\sum_{n=0}^\infty {\Gamma(\sigma-3/2) \sin(\eta\sqrt{\lambda_n})
\over \lambda_n \Gamma(1+2\sigma)\mathscr{I}_n}
\, K_n(x) \, Y_n(z)
\label{eq64}
\ ,
\end{equation}
where $\sigma$ is computed using Equation~(\ref{eq48}), and $Y_n$ and $K_n$
are defined in Equation~(\ref{eq52}). This is the same result as Equation~(27)
from KB, once we make the identifications $\tau_*=\eta$ and $G_n(\tau)=Y_n(z)$,
which arise due to the change in the spatial variable from the dimensionless
radius $z$ used here, to the scattering optical depth $\tau=\eta z$ used by KB.
The time-averaged X-ray spectrum computed using Equation~(\ref{eq64}) is compared
with the observational data for Cyg~X-1 and GX~339-04 in Section~6.1.
In Section~6.1.2, we use asymptotic analysis to derive a power-law
approximation to the exact radiation distribution given by Equation~(\ref{eq64}),
and we show that the resulting approximate X-ray spectrum agrees closely
with that obtained using the exact solution.

\subsection{Fourier Transform for $\alpha=0$}

In the homogeneous case ($\alpha=0$), we can substitute for the Fourier
transform $\Green$ in Equation~(\ref{eq36}) using the separation functions
\begin{equation}
F_{\lambda} \equiv H(\lambda,x) \, Y(\lambda,z)
\label{eq65}
\ ,
\end{equation}
to obtain, for $x \ne x_0$,
\begin{equation}
-\dfrac{1}{Y}\dfrac{1}{\eta^2 z^2}\dfrac{d}{dz}
\left(z^2 \dfrac{dY}{dz}\right)
=\dfrac{3\Theta}{H x^2} \dfrac{d}{dx}
\left[x^4\left(H + \dfrac{dH}{dx}\right)\right] + 3i\tilde{\omega}
= \lambda
\ ,
\label{eq66}
\end{equation}
where $\lambda=$constant. This relation can be broken into two ordinary
differential equations satisfied by the spatial and energy functions $Y$
and $H$. We obtain
\begin{equation}
\dfrac{1}{z^2} \dfrac{d}{dz}
\left(z^2 \dfrac{dY}{dz}\right) + \lambda \, \eta^2 Y = 0
\ ,
\label{eq67}
\end{equation}
\begin{equation}
{1 \over x^2}\dfrac{d}{dx}\left[x^4\left(H + \dfrac{dH}{dx}\right)
\right] - \dfrac{s}{3\Theta} \, H = 0
\ ,
\label{eq68}
\end{equation}
where
\begin{equation}
s \equiv \lambda-3i\tilde{\omega}
\ .
\label{eq69}
\end{equation}

In the Fourier transform case under consideration here, the spatial function
$Y$ must satisfy the mirror condition at the origin (cf. Equation~(\ref{eq49})),
\begin{equation}
\lim_{z \rightarrow 0} \ z^2 \, {dY(\lambda,z) \over dz} = 0
\label{eq69b}
\ .
\end{equation}
Since Equation~(\ref{eq67}) is identical to Equation~(\ref{eq44}), which
we previously encountered in Section~3.1 in our consideration of the
time-averaged spectrum produced in a homogeneous spherical corona,
we conclude that the fundamental solution for $Y$ is likewise given
by (cf. Equation~(\ref{eq50}))
\begin{equation} 
Y(\lambda,z) =
{\sin(\eta z\sqrt{\lambda}) \over \eta z}
\label{eq70}
\ .
\end{equation}
Furthermore, $Y$ must also satisfy the outer free-streaming boundary
condition, and therefore the eigenvalues $\lambda_n$ are the roots of
the equation (cf. Equation~(\ref{eq51}))
\begin{equation}
\lim_{z \rightarrow 1} \left[{1 \over 3\eta} {dY(\lambda,z) \over dz}
+ Y(\lambda,z)\right] = 0
\label{eq70b}
\ .
\end{equation}
{\it It follows that in a homogeneous corona, the Fourier eigenvalues $\lambda_n$
and spatial eigenfunctions $Y_n$ are exactly the same as those obtained in the
treatment of the time-averaged spectrum.} Hence we can also conclude that the spatial
eigenfunctions $Y_n$ form an orthogonal set, which motivates the development
of a series expansion for the Fourier transformed radiation Green's function, $\Green$.

Comparison of Equations~(\ref{eq68}) and (\ref{eq45}) allows us to
immediately obtain the solution for the energy function $H$ as (cf.
Equation~(\ref{eq46}))
\begin{equation}
H(\lambda,x) = (x x_0)^{-2} e^{-(x+x_0)/2} \, M_{2,\mu}(\xmin) \,
W_{2,\mu}(\xmax)
\ ,
\label{eq71}
\end{equation}
where $x_{\rm max}$ and $x_{\rm min}$ are defined in
Equations~(\ref{eq47}), and
\begin{equation}
\mu \equiv \sqrt{{9 \over 4} + {s \over 3\Theta}}
= \sqrt{{9 \over 4} + {\lambda-3i\tilde\omega \over 3\Theta}}
\ .
\label{eq72}
\end{equation}
Following the same steps used in Section~3.1 for the development of the
solution for the time-averaged radiation Green's function $\sgreen$, we can construct
a series representation for the Fourier transform $\Green$ by writing
\begin{equation}
\Green(x,z,\tilde\omega) = \sum_{n=0}^\infty
a_n \, H_n(x) \, Y_n(z)
\ ,
\label{eq73}
\end{equation}
where the eigenfunctions $Y_n$ and $H_n$ are defined by
\begin{equation}
Y_n(z) \equiv Y(\lambda_n,z) \ , \ \ \ 
H_n(x) \equiv H(\lambda_n,x)
\label{eq74}
\ .
\end{equation}
To solve for the expansion coefficients, $a_n$, we substitute
Equation~(\ref{eq73}) into Equation~(\ref{eq37}) with
$\alpha=0$ to obtain
\begin{equation}
\lim_{\delta \rightarrow 0}\sum_{n=0}^\infty
a_n Y_n(z)[H'(x_0+\delta)-H'(x_0-\delta)] = -\dfrac{N_0\,\delta(z-z_0)
e^{i \tilde{\omega} p_0}}{4\pi z_0^2 x_0^4 \Theta^4 (m_e c^2)^3 R^3}
\ ,
\label{eq75}
\end{equation}
or, equivalently,
\begin{equation}
\sum_{n=0}^\infty
a_n \, Y_n(z) \, \mathscr{W}_{2,\mu}(x_0)
= -\dfrac{N_0\,\delta(z-z_0)e^{i \tilde{\omega}
p_0}e^{x_0}}{4\pi z_0^2 \Theta^4 (m_e c^2)^3 R^3}
\label{eq76}
\ ,
\end{equation}
where the Wronskian is given by
\begin{equation}
\mathscr{W}_{2,\mu}(x_0) \equiv M_{2,\mu}(x_0)W'_{2,\mu}(x_0)
- W_{2,\mu}(x_0)M'_{2,\mu}(x_0) 
= -\dfrac{\Gamma(1+2\mu)}{\Gamma(\mu-3/2)}
\label{eq77}
\ .
\end{equation}
Substituting for the Wronskian in Equation~(\ref{eq76}) using
Equation~(\ref{eq77}) and applying the operator $\int_0^1 \eta^3
z^2 Y_m(z) dz$ to both sides of the equation, we can utilize the
orthogonality of the spatial eigenfunctions $Y_n$ to obtain for the
expansion coefficients $a_n$ the result
\begin{equation}
a_n = \dfrac{N_0\,e^{i \tilde{\omega} p_0}e^{x_0} \eta^3 \Gamma(\mu-3/2)
Y_n(z_0)} {4\pi\Theta^4 (m_e c^2)^3 R^3 \Gamma(1+2\mu)\mathscr{I}_n}
\label{eq78}
\ ,
\end{equation}
where the quadratic normalization integrals $\mathscr{I}_n$ are defined
in Equation~(\ref{eq54}).

By combining Equations~(\ref{eq73}) and (\ref{eq78}), we find
that the exact solution for the Fourier transformed radiation Green's function,
$\Green$, is given by the expansion
\begin{equation}
\Green(x,z,\tilde\omega) = {N_0\,e^{i \tilde{\omega} p_0} e^{x_0}
\eta^3 \over 4\pi R^3 \Theta^4 (m_e c^2)^3}
\sum_{n=0}^\infty
\dfrac{\Gamma(\mu-3/2)}
{\Gamma(1+2\mu)\mathscr{I}_n}
\, Y_n(z_0) Y_n(z) H_n(x)
\label{eq79}
\ ,
\end{equation}
with $\mu$ computed using Equation~(\ref{eq72}), and $Y_n$ and $H_n$ given by
Equations~(\ref{eq74}). This result agrees with Equation~(16) from KB once we
note the change in the spatial variable from $z$ to $\tau=\eta z$, with
$G_n(\tau)=Y_n(z)$, $\tau_*=\eta$, and $\ell_0=R/\eta$. In the case of the
exact solution for the time-averaged electron distribution derived in Section~3.1,
we are able to derive an accurate approximation using asymptotic analysis
(see Section~6.1.2). However, due to the complex nature of the series in
Equation~(\ref{eq79}), it is not possible to extract useful asymptotic
representations for the Fourier transform. Hence Equation~(\ref{eq79}) is the key
result that will be utilized to compute the Fourier transform and the
associated time lags for a spherical homogeneous cloud in Section~6.

\section{Inhomogeneous Model}

In the previous section, we have presented detailed solutions for the
time-averaged spectrum and for the Fourier transform of the time-dependent
photon Green's function describing the diffusion and Comptonization of photons
in a spherical, homogeneous scattering cloud. Another interesting
possibility is a coronal cloud with an electron number density
distribution that varies as $n_e(r) \propto r^{-1}$, which was considered
by HKC, and corresponds to $\alpha=1$ in Equations~(\ref{eq15}). In this
case, the dimensionless radius $z$ is related to the scattering optical
depth $\tau$ via (see Equations~(\ref{eq25}) and (\ref{eq26}))
\begin{equation}
\tau(z) = \eta \, \ln (z/z_{\rm in}) \ , \ \ \ \ \
\tau_* = \eta \, \ln (1/z_{\rm in})
\ ,
\label{eq80}
\end{equation}
where $\tau_*$ is the optical thickness measured from the inner radius
$r=r_{\rm in}$ ($z=z_{\rm in}$) to the outer radius $r=R$ ($z=1$). In
this section, we obtain the analytical solutions for the time-averaged
spectrum $\sgreen$ and for the Fourier transform $\Green$ for the case
with $n_e(r) \propto r^{-1}$.

\subsection{Quiescent Spectrum for $\alpha=1$}

The steady-state transport equation~(\ref{eq30}) describes the
formation of the time-averaged X-ray spectrum via the thermal Comptonization
of seed photons continually injected throughout a scattering corona
with an electron number density profile given by $n_e(r) \propto r^{-\alpha}$.
In the inhomogeneous case with $\alpha=1$, this equation can be solved
using the separation form
\begin{equation}
f_{\lambda} = K(\lambda,x) \ y(\lambda,z)
\label{eq81}
\ ,
\end{equation}
to obtain, for $x \ne x_0$,
\begin{equation}
\dfrac{-1}{y \eta^2 z}\dfrac{d}{dz}\left(z^3 {dy \over dz}\right)
= {3 \, \Theta \over K \, x^2}{d \over dx}
\left[x^4\left(K + {dK \over dx}\right)\right] = \lambda
\label{eq82}
\ ,
\end{equation}
where $\lambda=$constant. The associated ordinary differential
equations in the spatial and energy coordinates are, respectively,
\begin{equation}
{1 \over z}\dfrac{d}{dz}\left(z^3 {dy \over dz}\right) + \lambda
\, \eta^2 y = 0
\label{eq83}
\ ,
\end{equation}
\begin{equation}
{1 \over x^2}\dfrac{d}{dx} \left[x^4\left(K + {dK \over dx}\right)
\right] - \dfrac{\lambda}{3\Theta} \, K = 0
\label{eq84}
\ .
\end{equation}
Since Equation~(\ref{eq84}) is identical to Equation~(\ref{eq45}),
it follows that the solution for the energy function $K$ is given by
(cf. Equation~(\ref{eq46}))
\begin{equation}
K(\lambda,x) = (x x_0)^{-2} e^{-(x+x_0)/2} M_{2,\sigma}(x_{\rm min})
W_{2,\sigma}(x_{\rm max})
\label{eq85}
\ ,
\end{equation}
where
\begin{equation}
\sigma \equiv \sqrt{\dfrac{9}{4}+\dfrac{\lambda}{3\Theta}}
\ .
\label{eq86}
\end{equation}

The fundamental solutions for the spatial functions, $y$, are
given by the power-law forms
\begin{equation}
y(\lambda,z) = C_1 z^{-1-\sqrt{1-\eta^2 \lambda}}
+ z^{-1+\sqrt{1-\eta^2 \lambda}}
\label{eq87}
\ ,
\end{equation}
where $C_1$ is a superposition constant determined by applying the outer
free-streaming boundary condition given by Equation~(\ref{eq39}). For
the inhomogeneous case with $\alpha=1$, the outer boundary condition
implies that $y$ must satisfy the equation
\begin{equation} 
\lim_{z \rightarrow 1} \left[{z \over 3\eta} {dy(\lambda,z) \over dz}
+ y(\lambda,z)\right] = 0
\label{eq88}
\ .
\end{equation}
The corresponding result obtained for $C_1$ is
\begin{equation} 
C_1 = {3\eta -1 + \sqrt{1-\eta^2 \lambda} \over
1 - 3\eta + \sqrt{1-\eta^2 \lambda}}
\label{eq89}
\ .
\end{equation}

The next step is to apply the {\it inner} free-streaming boundary
condition, given by Equation~(\ref{eq40}). Stated in terms of
$z$, we obtain for $\alpha=1$ the condition
\begin{equation} 
\lim_{z \rightarrow z_{\rm in}} \left[{z \over 3\eta} {dy(\lambda,z) \over dz}
- y(\lambda,z)\right] = 0
\label{eq90}
\ ,
\end{equation}
where $z_{\rm in}=r_{\rm in}/R$ is the dimensionless inner radius of the
cloud. Equation~(\ref{eq90}) is satisfied only for certain discrete
values of $\lambda$, which are the eigenvalues $\lambda_n$. The
eigenvalues obtained are all positive real numbers. The resulting global
functions $y$ therefore satisfy both the inner and outer free-streaming
boundary conditions. Once the eigenvalues $\lambda_n$ are determined,
the corresponding spatial and energy eigenfunctions are defined by
\begin{equation}
y_n(z) \equiv y(\lambda_n,z) \ , \ \ \
K_n(x) \equiv K(\lambda_n,x)
\label{eq91}
\ .
\end{equation}

We show in Appendix~A that the spatial eigenfunctions $y_n$ form an
orthogonal set with respect to the weight function $z$, so that
\begin{equation}
\int_{z_{\rm in}}^1 z \, y_n(z) \, y_m(z) \, dz = 0 \ , \ \ \ \ \ n \ne m
\label{eq92}
\ .
\end{equation}
We can therefore express the steady-state photon Green's function $\sgreen$
using the expansion
\begin{equation}
\sgreen(x,x_0,z) = \sum_{n=0}^{\infty} c_n \, K_n(x) \, y_n(z)
\label{eq93}
\ .
\end{equation}
To solve for the expansion coefficients, $c_n$, we substitute
Equation~(\ref{eq93}) into Equation~(\ref{eq32}), with
$\alpha=1$, to obtain
\begin{equation}
\lim_{\delta \rightarrow 0}\sum_{n=0}^\infty
c_n y_n(z)[K_n'(x_0+\delta)-K_n'(x_0-\delta)]
= -\dfrac{\dot N_0}{2\pi R^2 \eta c \, \Theta^4 (m_e c)^3 x_0^4
(1-z_{\rm in}^2)}
\ .
\label{eq94}
\end{equation}
Eliminating $K$ using Equation~(\ref{eq46}) yields
\begin{equation}
\sum_{n=0}^\infty
c_n y_n(z) \, \mathscr{W}_{2,\sigma}(x_0)
= -\dfrac{\dot N_0 e^{x_0}}{2\pi R^2 \eta c \, \Theta^4 (m_e c)^3
(1-z_{\rm in}^2)}
\label{eq95}
\ ,
\end{equation}
where the Wronskian $\mathscr{W}_{2,\sigma}(x_0)$ is defined in
Equation~(\ref{eq59}). By combining Equations~(\ref{eq95})
and (\ref{eq60}) we obtain
\begin{equation}
\sum_{n=0}^{\infty} c_n y_n(z) \, {\Gamma(1+2\sigma)
\over \Gamma(\sigma-3/2)}
= \dfrac{\dot N_0 e^{x_0}}{2\pi R^2 \eta c \, \Theta^4 (m_e c)^3
(1-z_{\rm in}^2)}
\ .
\label{eq96}
\end{equation}

We can exploit the orthogonality of the spatial basis functions
$y_n(z)$ with respect to the weight function $z$ by operating on
Equation~(\ref{eq96}) with $\int_{z_{\rm in}}^1 z \, y_m(z)dz$
to obtain
\begin{equation}
c_n = \dfrac{\dot N_0 e^{x_0}\Gamma(\sigma-3/2)\mathscr{L}_n}
{2\pi R^2 \eta c \, \Theta^4 (m_e c^2)^3 \mathscr{J}_n \Gamma(1+2\sigma)
(1-z_{\rm in}^2)}
\label{eq97}
\ ,
\end{equation}
where we have made the definitions
\begin{equation}
\mathscr{J}_n \equiv \int_{z_{\rm in}}^1 z \, y_n^2(z) dz \ , \ \ \ \ \
\mathscr{L}_n \equiv \int_{z_{\rm in}}^1 z \, y_n(z) dz
\label{eq98}
\ .
\end{equation}
The final result for the steady-state (quiescent) photon Green's function
in the inhomogeneous case with $\alpha=1$ is obtained by combining
Equations~(\ref{eq93}) and (\ref{eq97}), which yields
\begin{equation}
\sgreen(x,x_0,z) = {\dot N_0 e^{x_0} \over 2\pi R^2 \eta c \, \Theta^4 (m_e c^2)^3}
\sum_{n=0}^\infty \dfrac{\Gamma(\sigma-3/2) \mathscr{L}_n}
{\mathscr{J}_n \Gamma(1+2\sigma)(1-z_{\rm in}^2)} \, K_n(x) \, y_n(z)
\label{eq99}
\ ,
\end{equation}
with $\sigma$ computed using Equation~(\ref{eq86}), and $y_n$ and $K_n$
given by Equations~(\ref{eq91}). The time-averaged X-ray spectrum computed
using Equation~(\ref{eq99}) is compared with observational data for two specific
sources in Section~6.1, and an accurate asymptotic approximation is derived
in Section~6.1.2.

\subsection{Fourier Transform for $\alpha=1$}

In the inhomogeneous case with $\alpha=1$, we can substitute for the
Fourier transform in Equation~(\ref{eq36}) using the separation
functions
\begin{equation}
F_{\lambda} \equiv K(\lambda,x) \ g(\lambda,z)
\label{eq100}
\ ,
\end{equation}
to obtain, for $x \ne x_0$,
\begin{equation}
-\dfrac{1}{g}\dfrac{1}{\eta^2 z}\dfrac{d}{dz}
\left(z^3 \dfrac{dg}{dz}\right) - 3i\tilde{\omega} z
= \dfrac{3\Theta}{K x^2} \dfrac{d}{dx}
\left[x^4\left(K + \dfrac{dK}{dx}\right)\right]
= \lambda
\ ,
\label{eq101}
\end{equation}
where $\lambda$ is the separation constant. This relation yields
two ordinary differential equations satisfied by the spatial and
energy functions $g$ and $K$, given by
\begin{equation}
\dfrac{1}{z}\dfrac{d}{dz}\left(z^3 \dfrac{dg}{dz}\right) +\eta^2
\Big(\lambda + 3i\tilde{\omega}z\Big) g = 0
\ ,
\label{eq102}
\end{equation}
\begin{equation}
\dfrac{1}{x^2}\dfrac{d}{dx}\left[x^4\left(K + \dfrac{dK}{dx}\right)
\right]-\dfrac{\lambda}{3\Theta} K = 0
\ .
\label{eq103}
\end{equation}
Equation~(\ref{eq103}) is identical to Equation~(\ref{eq45}),
and therefore we can immediately conclude that the solution for the
energy function $K$ is given by
\begin{equation}
K(\lambda,x) = (x x_0)^{-2} e^{-(x+x_0)/2} M_{2,\sigma}(x_{\rm min})
W_{2,\sigma}(x_{\rm max})
\label{eq104}
\ ,
\end{equation}
where
\begin{equation}
\sigma \equiv \sqrt{\dfrac{9}{4}+\dfrac{\lambda}{3\Theta}}
\ .
\label{eq105}
\end{equation}

One significant new feature in the inhomogeneous case with $\alpha=1$
under consideration here is that the eigenvalues $\lambda_n$ are now
functions of the Fourier frequency $\tilde\omega$, which stems from
the appearance of $\tilde\omega$ in Equation~(\ref{eq102}).
It follows that $\sigma$ is also a function of $\tilde\omega$ through
its dependence on $\lambda$ (see Equation~(\ref{eq48})). This
inconvenient mixing of variables forces us to generate a separate list
of eigenvalues for each sampled frequency. The fundamental solution
for the spatial function $g$ is given by the superposition
\begin{equation}
g(\lambda,z) = \dfrac{1}{z}\left[
C_2 J_{-\nu}(2\eta \sqrt{3i\tilde{\omega}z})
+ J_\nu(2\eta\sqrt{3i\tilde{\omega}z})\right]
\label{eq106}
\ ,
\end{equation}
where $J_\nu(z)$ denotes the Bessel function of the first kind,
and we have made the definition
\begin{equation}
\nu \equiv 2\sqrt{1-\eta^2 \lambda}
\label{eq107}
\ .
\end{equation}
The superposition constant $C_2$ is computed by applying the outer
free-streaming boundary condition, which can be written as (cf.
Equation~(\ref{eq88}))
\begin{equation} 
\lim_{z \rightarrow 1} \left[{z \over 3\eta} {dg(\lambda,z) \over dz}
+ g(\lambda,z)\right] = 0
\label{eq108}
\ .
\end{equation}
The result obtained for $C_2$ is
\begin{equation} 
C_2 = {(2 - 6\eta + \nu)J_\nu(2\eta\sqrt{3i\tilde\omega})
- 2 \eta \sqrt{3i\tilde\omega}J_{\nu-1}(2\eta\sqrt{3i\tilde\omega})
\over (6\eta - 2 + \nu)J_{-\nu}(2\eta\sqrt{3i\tilde\omega})
+ 2 \eta \sqrt{3i\tilde\omega}J_{-\nu-1}(2\eta\sqrt{3i\tilde\omega})}
\label{eq109}
\ .
\end{equation}

Next we must apply the inner free-streaming boundary condition
given by (cf. Equation~(\ref{eq90}))
\begin{equation} 
\lim_{z \rightarrow z_{\rm in}} \left[{z \over 3\eta} {dg(\lambda,z) \over dz}
- g(\lambda,z)\right] = 0
\label{eq110}
\ ,
\end{equation}
where $z_{\rm in}=r_{\rm in}/R$. The roots of Equation~(\ref{eq110})
are the eigenvalues $\lambda_n$, and the associated global functions
$g$ satisfy both the inner and outer free-streaming boundary conditions.
The corresponding spatial and energy eigenfunctions are given by
\begin{equation}
g_n(z) \equiv g(\lambda_n,z) \ , \ \ \
K_n(x) \equiv K(\lambda_n,x)
\label{eq111}
\ .
\end{equation}

As demonstrated in Appendix~A, the spatial eigenfunctions $g_n$ are
orthogonal with respect to the weight function $z$, and therefore
\begin{equation}
\int_{z_{\rm in}}^1 z \, g_n(z) \, g_m(z) \, dz = 0 \ , \ \ \ \ \ n \ne m
\label{eq112}
\ .
\end{equation}
It follows that we can express the Fourier transformed radiation Green's
function, $\Green$, using the expansion (cf. Equation~(\ref{eq73}))
\begin{equation}
\Green(x,z,\omega) = \sum_{n=0}^{\infty} d_n \, K_n(x) \, g_n(z)
\label{eq113}
\ .
\end{equation}
The expansion coefficients $d_n$ can be computed by applying the
derivative jump condition given by Equation~(\ref{eq37}), which
yields for $\alpha=1$
\begin{equation}
\lim_{\delta \rightarrow 0}\left[\dfrac{d\Green}{dx}
\right]\Bigg|_{x_0-\delta}^{x_0+\delta} = -\dfrac{N_0\,\delta(z-z_0)
e^{i \tilde{\omega} p_0}}{4\pi x_0^4 \, z_0^2 z^{-1} \, \Theta^4
(m_e c^2)^3 R^3}
\ .
\label{eq114}
\end{equation}
Combining Equations~(\ref{eq113}) and (\ref{eq114}) gives the result
\begin{equation}
\lim_{\delta \rightarrow 0}\sum_{n=0}^{\infty}
d_n g_n(z)[K'(x_0+\delta)-K'(x_0-\delta)] = -\dfrac{N_0\,\delta(z-z_0)
e^{i \tilde{\omega} p_0}}{4\pi x_0^4 \, z_0^2 z^{-1} \, \Theta^4
(m_e c^2)^3 R^3}
\label{eq115}
\ ,
\end{equation}
or, equivalently,
\begin{equation}
\sum_{n=0}^\infty
d_n g_n(z) \, \mathscr{W}_{2,\sigma}(x_0)
= \sum_{n=0}^{\infty} - d_n g_n(z) \, {\Gamma(1+2\sigma)
\over \Gamma(\sigma-3/2)}
= -\dfrac{N_0\,\delta(z-z_0)
e^{i \tilde{\omega} p_0}e^{x_0}}{4\pi z_0^2 z^{-1} \, \Theta^4
(m_e c^2)^3 R^3}
\label{eq116}
\ ,
\end{equation}
where we have utilized Equations~(\ref{eq59}) and (\ref{eq60})
for the Wronskian $\mathscr{W}_{2,\sigma}(x_0)$.

We can solve for the expansion coefficients $d_n$ by utilizing the
orthogonality of the spatial eigenfunctions $g_n$ with respect to the
weight function $z$. Applying $\int_{z_{\rm in}}^{1}z g_m(z)dz$ to both
sides of Equation~(\ref{eq116}), we obtain, after some algebra,
\begin{equation}
d_n = \dfrac{N_0 e^{i\tilde{\omega}p_0}e^{x_0}\Gamma(\sigma-3/2)g_n(z_0)}
{4\pi \Theta^4 (m_e c^2)^3 R^3 \Gamma(1+2\sigma) \mathscr{K}_n}
\ ,
\label{eq117}
\end{equation}
where the quadratic normalization integrals, $\mathscr{K}_n$, are
defined by
\begin{equation}
\mathscr{K}_n \equiv \int_{z_{\rm in}}^1 z g^2_n(z)dz
\label{eq118}
\ .
\end{equation}
The final result for the Fourier transform $\Green$ of the photon Green's
function $\green$ obtained by combining Equations~(\ref{eq113})
and (\ref{eq117}) is
\begin{equation}
\Green(x,z,\tilde\omega) = \dfrac{N_0 \, e^{i
\tilde{\omega}p_0}e^{x_0}}{4\pi R^3 \Theta^4 (m_e c^2)^3}
\sum_{n=0}^\infty \dfrac{\Gamma(\sigma-3/2)}{\Gamma(1+2\sigma)
\mathscr{K}_n} \, g_n(z_0) \, g_n(z) \, K_n(x)
\label{eq119}
\ ,
\end{equation}
with $\sigma$ evaluated using Equation~(\ref{eq105}), and $g_n$ and $K_n$
given by Equations~(\ref{eq111}). This exact analytical solution can be used
to generate theoretical predictions of the Fourier transformed data streams
in two different energy channels in order to simulate the time lags created
in a spherical scattering corona with an electron number density profile that
varies as $n_e(r) \propto r^{-1}$. As in the case of the homogeneous Fourier
transform discussed in Section~3.2, it is not possible to extract useful asymptotic
representations for the inhomogeneous Fourier transform due to the complex
nature of the sum appearing in Equation~(\ref{eq119}).

\section{Bremsstrahlung Injection}

The investigations carried out by Miyamoto (1988), HKC, and KB show that
the impulsive injection of monochromatic seed photons into a homogeneous
Comptonizing corona cannot produce the observed dependence of the X-ray
time lags on the Fourier frequency. A major advantage of the analytical
method we employ here is that {\it the radiation Green's function we obtain can be
convolved with any desired seed photon distribution as a function of
radius $r$, energy $\epsilon$, and time $t$.} This flexibility stems
from the fact that the transport equation is a linear partial
differential equation. A source spectrum of particular interest is a
flash of bremsstrahlung seed photons injected on a spherical shell at
radius $r=r_0$. We may expect the observed variability in this case to
be qualitatively different from the behavior associated with a
monochromatic flash of seed photons, because the bremsstrahlung flash
represents {\it broadband} radiation. We anticipate that the prompt
escape of high-energy photons from the bremsstrahlung seed distribution
may cause a profound shift in the dependence of the observed X-ray time
lags on the Fourier period.

Since the fundamental transport equation governing the radiation field
is linear, it follows that we can compute the time-dependent spectrum
$f$ resulting from any seed photon distribution $Q$ that is an arbitrary
function of time, energy, and radius using the integral convolution
\begin{equation}
f(\epsilon,r,t) = \int_0^\infty\int_{r_{\rm in}}^R
\int_0^\infty 4 \pi r_0^2 \epsilon_0^2
\green(\epsilon,\epsilon_0,r,r_0,t,t_0)
\, Q(\epsilon_0,r_0,t_0) N_0^{-1} \, d\epsilon_0 \, dr_0
\, dt_0
\ ,
\label{eq120}
\end{equation}
where $4\pi r_0^2 \epsilon_0^2 Q(\epsilon_0,r_0,t_0) \, dr_0 \, dt_0 \,
d\epsilon_0$ gives the number of photons injected in the energy range
$d\epsilon_0$, radius range $dr_0$, and time range $dt_0$ around the
coordinates $(\epsilon_0,r_0,t_0)$. In the case of optically thin
bremsstrahlung injection, the seed photons are created as a result of a
local instability in the coronal plasma, due to, for example, a magnetic
reconnection event, or the passage of a shock. It follows that the photon
distribution resulting from localized, impulsive injection of bremsstrahlung
radiation at radius $r=r_0$ can be written as
\begin{equation}
f_{\rm brem}(\epsilon,r,t) = \int_{\epsilon_{\rm abs}}^\infty
\green(\epsilon,\epsilon_0,r,r_0,t,t_0)
\, Q_{\rm brem}(\epsilon_0) N_0^{-1} \, d\epsilon_0
\ ,
\label{eq121}
\end{equation}
where $\epsilon_{\rm abs}$ denotes the low-energy cutoff due to free-free
self-absorption in the source plasma, and the bremsstrahlung source function,
$Q_{\rm brem}$, for fully-ionized hydrogen is given by (Rybicki \& Lightman 1979)
\begin{equation}
Q_{\rm brem}(\epsilon_0) = {A_0 \over \epsilon_0}
\, e^{-\epsilon_0/k T_e}
\ ,
\label{eq122}
\end{equation}
where
\begin{equation}
A_0 = {2^5 \pi q^6 \over 3 h m_e c^3}
\left(2 \pi \over 3 k m_e \right)^{1/2}
V_0 \, t_{\rm rad} \,
T_e^{-1/2} n_e^2(r_0)
\ .
\label{eq122b}
\end{equation}
Here, $V_0$ denotes the radiating volume, $t_{\rm rad}$ is the radiating
time interval, and $q$ is the electron charge. The bremsstrahlung source
function is normalized so that $Q_{\rm brem}(\epsilon_0) \, d\epsilon_0$
gives the number of photons injected in the energy range between
$\epsilon_0$ and $\epsilon_0+d\epsilon_0$.

The low-energy self-absorption cutoff, $\epsilon_{\rm abs}$, appearing
in Equation~(\ref{eq121}), depends on the temperature and density of the
plasma experiencing the transient that produces the flash of
bremsstrahlung seed photons. The density of the unstable plasma
is expected to be higher than that in the surrounding
corona, due to either shock compression or a thermal instability. We do not
analyze this physical process in detail here, and instead we treat $\epsilon_{\rm
abs}$ as a free parameter in our model, although a more detailed
physical picture could be developed in future work.

Changing variables from $(\epsilon,r,t)$ to $(x,z,p)$ and applying
Fourier transformation to both sides of Equation~(\ref{eq121}), we
obtain
\begin{equation}
F_{\rm brem}(x,z,z_0,\tilde\omega) = A_0 \, N_0^{-1} \int_{x_{\rm abs}}^\infty
\Green(x,x_0,z,z_0,\tilde\omega)
\, x_0^{-1} e^{-x_0} \, dx_0
\ ,
\label{eq123}
\end{equation}
where $x_{\rm abs} = \epsilon_{\rm abs}/(k T_e)$ is the
dimensionless self-absorption energy. The function $\Green$ in
Equation~(\ref{eq123}) represents the Fourier transformation of the
time-dependent photon Green's function for either the homogeneous or
inhomogeneous cases, given by either Equation~(\ref{eq79}) or
Equation~(\ref{eq119}), respectively. The integral with respect to $x_0$ can
be carried out analytically, and the exact solutions are given by
\begin{equation}
F_{\rm brem}(x,z,z_0,\tilde\omega)=
\dfrac{e^{i \tilde{\omega} p_0} \eta^3 A_0 \, e^{-x/2}}{4 \pi R^3
\Theta^4 (m_e c^2)^{3} x^2}\sum_{n=0}^\infty
\begin{cases}
\dfrac{\Gamma(\mu-3/2) Y_n(z_0) Y_n(z)}{\Gamma(1+2\mu)
\mathscr{I}_n} \, B(\mu,x) \, , & {\rm homogeneous} \, , \\ 
\dfrac{\Gamma(\sigma-3/2)g_n(z_0) g_n(z)}{\Gamma(1+2\sigma)
\, \eta^3 \mathscr{K}_n} \,  B(\sigma,x) \, , & {\rm inhomogeneous} \, ,
\label{eq124}
\end{cases}
\end{equation}
where $\sigma$ and $\mu$ are given by Equations~(\ref{eq48}) and
(\ref{eq72}), respectively, and the integral function $B(\lambda,x)$ is
defined by
\begin{equation}
B(\lambda,x) \equiv \int_{x_{\rm abs}}^{\infty} e^{-x_0/2}
x_0^{-3} M_{2,\lambda}(x_{\rm min})W_{2,\lambda}(x_{\rm max})
\, dx_0
\label{eq125}
\ .
\end{equation}
We show in Appendix~B that $B(\lambda,x)$ can be evaluated analytically
to obtain the closed-form result
\begin{equation}
B(\lambda,x) =
\begin{cases}
W_{2,\lambda}(x) [I_M(\lambda,x) - I_M(\lambda,x_{\rm abs})] - M_{2,\lambda}(x)
I_W(\lambda,x) \, , & x \ge x_{\rm abs} \, , \\
- M_{2,\lambda}(x) I_W(\lambda,x_{\rm abs}) \, , & x \le x_{\rm abs} \, ,
\end{cases}
\label{eq126}
\end{equation}
where the functions $I_M$ and $I_W$ are defined by
\begin{equation}
I_M(\lambda,x) \equiv \dfrac{x^{-2} e^{-x/2}}{\lambda+\frac{3}{2}}
\bigg(M_{1,\lambda}(x)
+ \dfrac{3}{\lambda+\frac{1}{2}} \bigg\{M_{0,\lambda}(x)
+ \dfrac{2}{\lambda-\frac{1}{2}} \bigg[M_{-1,\lambda}(x)
+ \dfrac{1}{\lambda-\frac{3}{2}} M_{-2,\lambda}(x) \bigg] \bigg\} \bigg)
\ ,
\label{eq127}
\end{equation}
and
\begin{equation}
I_W(\lambda,x) \equiv x^{-2} e^{-x/2} \bigg[-W_{1,\lambda}(x) + 3 W_{0,\lambda}(x)
- 6 W_{-1,\lambda}(x) + 6 W_{-2,\lambda}(x) \bigg]
\label{eq128}
\ .
\end{equation}
Section~6, we will use this result to study the
implications of broadband (bremsstrahlung) seed photon injection as an
alternative to monochromatic injection for the production of the
observed X-ray time lags in homogeneous and inhomogeneous scattering
coronae.

\section{Astrophysical Applications}

In the previous sections, we have obtained the exact mathematical
solution for the steady-state photon Green's function, $\sgreen$, describing
the X-ray emission emerging from a scattering corona as a result of the {\it
continual distributed} injection of monochromatic seed photons.
We have also obtained the exact solution for the
Green's function, $\Green$, describing the Fourier transform of the
X-ray spectrum resulting from the {\it impulsive localized} injection of
monochromatic seed photons into the corona. By convolving the solution for
$\Green$ with the bremsstrahlung source term, we were also able to derive the
exact solution for the bremsstrahlung Fourier transform, $F_{\rm brem}$.

The availability of these various solutions for the steady-state X-ray spectrum
and for the Fourier transform resulting from impulsive injection allows us to
explore a wide variety of injection scenarios, while maintaining explicit control
over the physical parameters describing the astrophysical objects of
interest, such as the temperature, the electron number density,
and the cloud radius. Our goal here is to develop ``integrated models,''
in which the coupled calculations of the time-averaged X-ray spectrum and
the transient Fourier X-ray time lags are based on the {\it same set} of
physical parameters (temperature, density, radius) for the scattering
corona. We believe that this integrated approach
represents a significant step forward by facilitating the study of
a broad range of parameter space using an analytical model.

\subsection{Comparison with Observed Time-Averaged Spectra}

The time-averaged X-ray
spectrum emanating from the outer surface of the cloud results from the
continual distributed injection of soft photons from a source with a rate
that is proportional to the local electron number density. Thus, there is
no specific injection radius for the time-averaged model. The detailed solutions
we have obtained describe the radiative transfer occurring in either a
homogeneous cloud, or in an inhomogeneous cloud in which the electron
number density varies with radius as $n_e(r) \propto r^{-1}$.

Application of the integrated model begins with a comparison of
the observed time-averaged X-ray spectrum with the theoretical steady-state
photon flux measured at the detector, ${\cal F}_\epsilon(\epsilon)$,
computed using the relation
\begin{equation}
{\cal F}_\epsilon(\epsilon) = \left(R \over D\right)^2 c \ \epsilon^2
\sgreen\left({\epsilon\over k T_e},x_0,z\right) \bigg|_{z=1}
\ ,
\label{eq129}
\end{equation}
where $D$ is the distance to the source, $R$ is the radius of the
corona, $c$ is the speed of light, and the solution for the steady-state
spectrum, $\sgreen(x,x_0,1)$, at the surface of the cloud is evaluated
using either Equation~(\ref{eq64}) for the homogeneous case or
Equation~(\ref{eq99}) for the inhomogeneous case. In our computations
of the time-averaged X-ray spectra, the seed photon energy is frozen at
$\epsilon_0=0.1$ keV in order to approximate the effect of the continual
injection of blackbody photons from a ``cool'' accretion disk with
temperature $T \sim 10^6\,$K.

The temperature parameter $\Theta = k T_e/m_e c^2$ (Equation~(\ref{eq18}))
and the scattering parameter $\eta = R/\ell_*$ (Equation~(\ref{eq20})) determine
the slope of the power-law component of the time-averaged spectrum, and also the
frequency of the high-energy exponential cutoff created by recoil losses.
In the inhomogeneous case, the shape of the time-averaged spectrum also depends
on the dimensionless inner radius, $z_{\rm in}=r_{\rm in}/R$, at which the inner
free-streaming boundary condition is imposed. We vary the values
of $\Theta$, $\eta$, and $z_{\rm in}$ until good qualitative agreement with
the shape of the observed steady-state X-ray spectrum is achieved.
Once the values of $\Theta$, $\eta$, and $z_{\rm in}$ are determined,
the photon injection rate, $\dot N_0$, is then computed by matching the
theoretical flux level with the observed time-averaged spectrum.

\subsubsection{Exact Time-Averaged X-ray Spectra}

In Figure~1, we plot the theoretical time-averaged (quiescent) X-ray spectra measured
at the detector, ${\cal F}_\epsilon(\epsilon)$, computed using the {\it homogenous}
corona model, with distributed seed photon injection, evaluated by combining
Equations~(\ref{eq64}) and (\ref{eq129}). The plots also include a comparison
with the observed X-ray spectra for Cyg~X-1 and GX~339-04. The data for Cyg X-1 were reported
by Cadolle Bel et al. (2006) and cover the observation period MJD 52617-52620,
and the data for GX 339-04 were reported by Cadolle Bel et al. (2011) and cover
the observation period MJD 55259.9-55261.1. Both sources were observed
by INTEGRAL in the low/hard state. The model parameters are summarized in
Table~1, and the corresponding homogeneous eigenvalues are plotted in
Figure~3. The time-averaged X-ray spectra obtained for the {\it inhomogeneous}
corona model, computed by combining Equations~(\ref{eq99}) and (\ref{eq129}),
are plotted and compared with the observational data in Figure~2, and the
corresponding inhomogeneous eigenvalues are depicted in Figure 3.

We find that the observed time-averaged spectra can be fit equally well using either
the homogeneous or inhomogeneous cloud models. Furthermore, the homogeneous
and inhomogeneous models have similar temperatures and cloud radii. This
behavior illustrates the fact that the time-averaged spectrum mainly depends
on the cloud temperature and the Compton $y$-parameter, and is not directly
dependent on the accretion geometry, as discussed in detail by Sunyaev \&
Titarchuk (1980).

\begin{figure}[h!]
\captionsetup[subfigure]{labelformat=empty}
\subfloat[]{\includegraphics[scale=0.85]{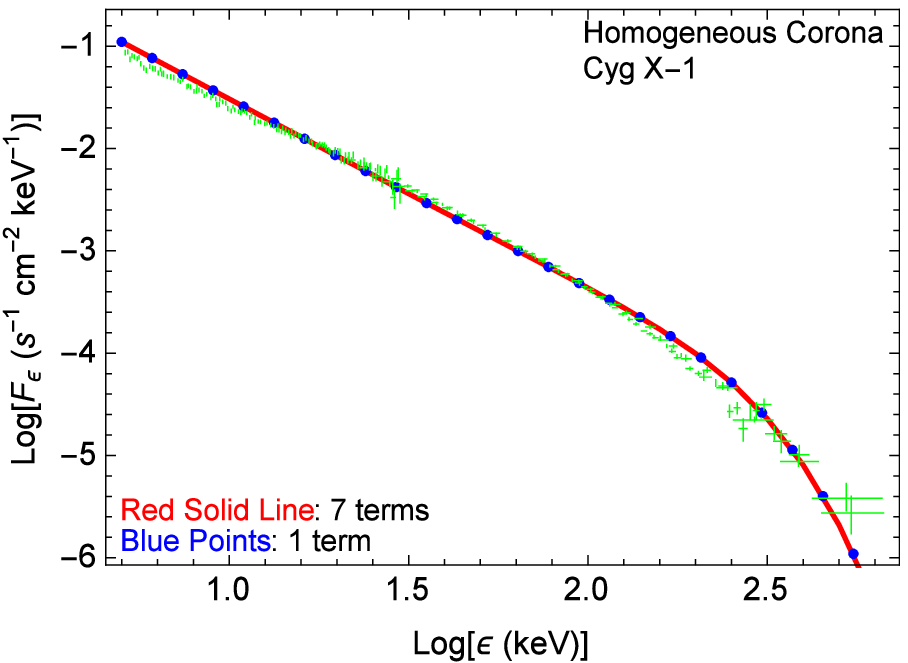}}
\subfloat[]{\includegraphics[scale=0.87]{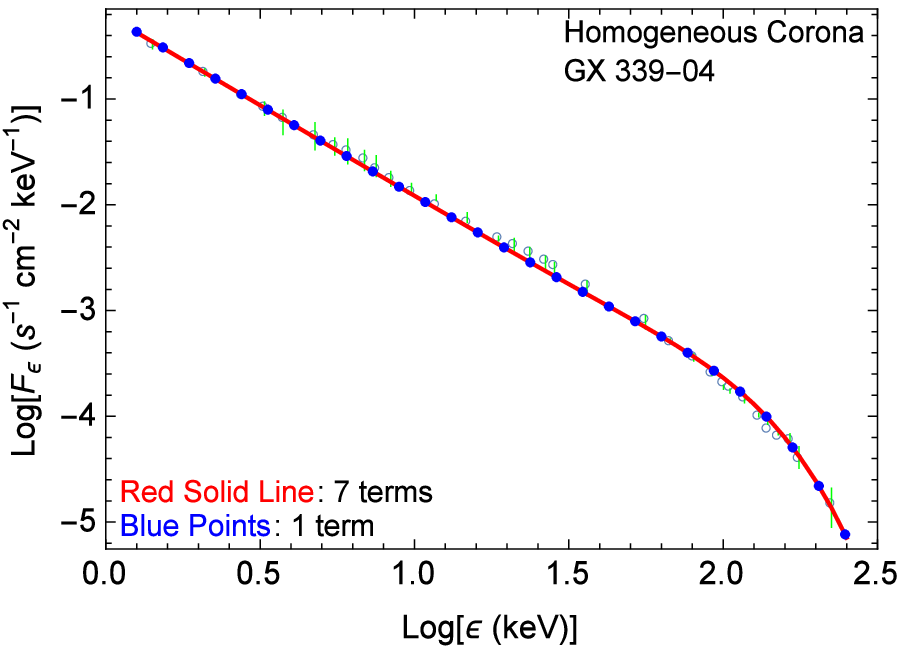}}
\caption{Theoretical time-averaged (quiescent) X-ray spectra, ${\cal F}_\epsilon(\epsilon)$, observed
at the detector, for a homogeneous corona, with constant electron number
density, $n_e$, computed by combining Equations~(\ref{eq64}) and (\ref{eq129}).
Results are presented for Cyg~X-1 (left panel) and GX~339-04 (right panel), along
with observational data taken from Cadolle Bel et al. (2006) and Cadolle Bel et al.
(2011), respectively. Both sources were observed in the low/hard
state using INTEGRAL. To analyze the convergence of the series, we plot
the results obtained using only the first term in the series, or using the first
7 terms. The convergence is extremely rapid for both sources.}
\end{figure}

\begin{figure}[h!]
\captionsetup[subfigure]{labelformat=empty}
\subfloat[]{\includegraphics[scale=0.85]{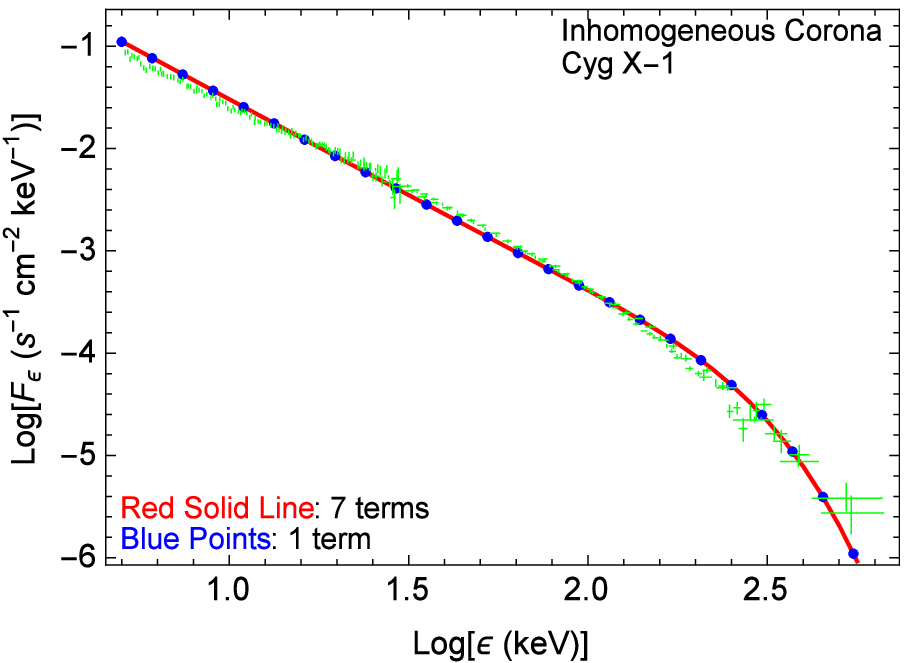}}
\subfloat[]{\includegraphics[scale=0.87]{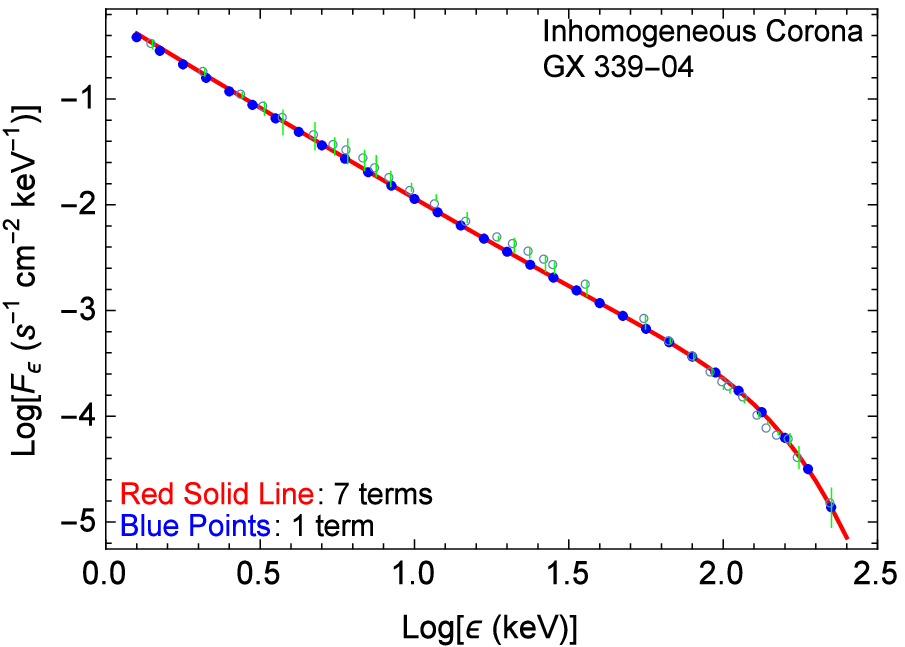}}
\caption{Same as Figure~1, except we plot the time-averaged X-ray
spectra emanating from an inhomogeneous corona, with electron density
profile $n_e(r) \propto r^{-1}$. The results were obtained by combining
Equations~(\ref{eq99}) and (\ref{eq129}). The convergence is very rapid.}
\end{figure}

\begin{figure}[h!]
\captionsetup[subfigure]{labelformat=empty}
\subfloat[]{\includegraphics[scale=0.88]{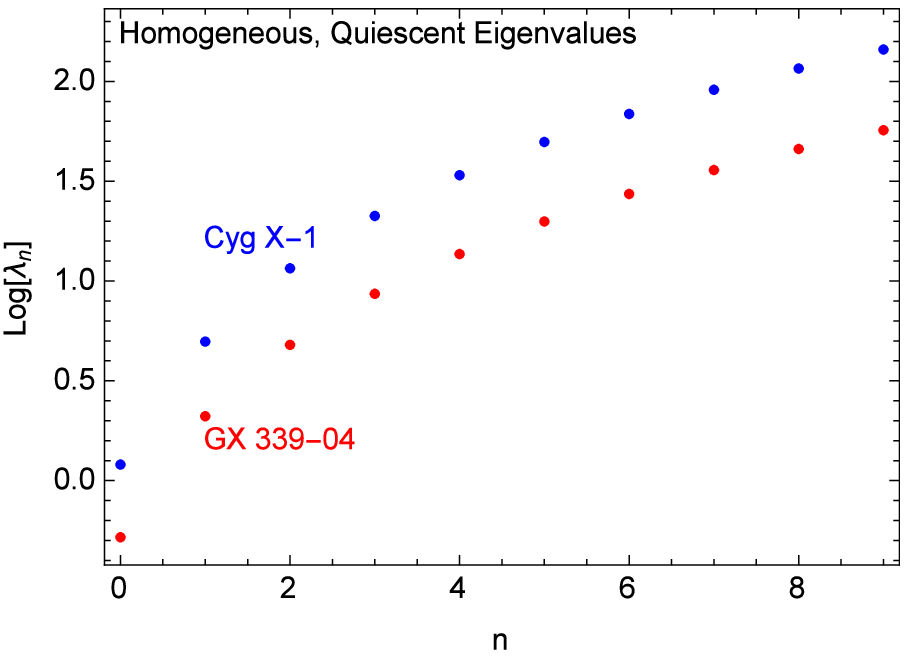}}
\subfloat[]{\includegraphics[scale=0.85]{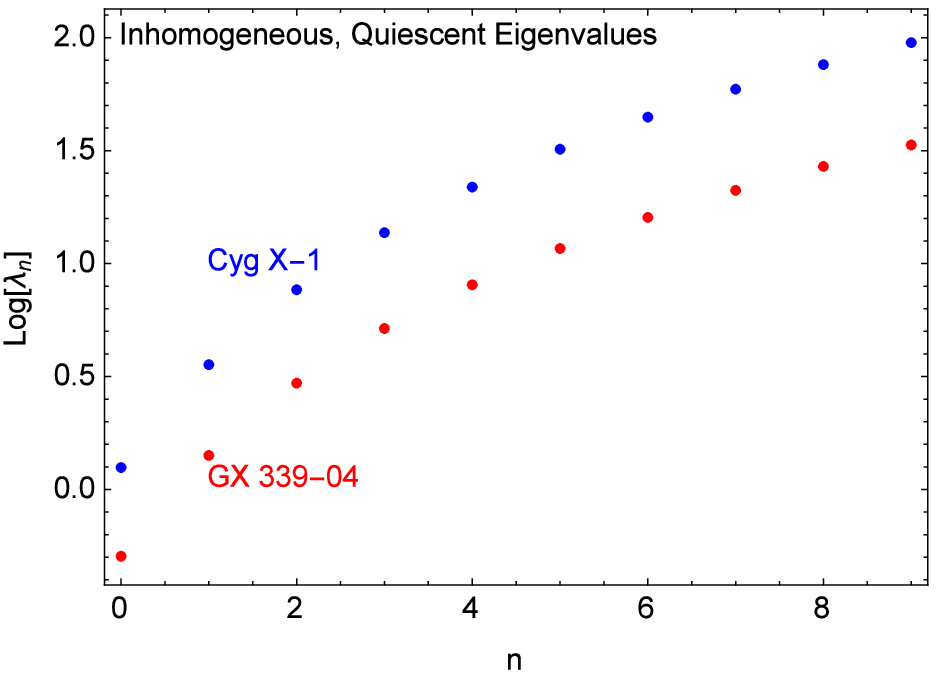}}
\caption{Real eigenvalues, $\lambda_n$, for the time-averaged (quiescent) spectrum radiated by
a homogeneous corona (left panel), and an inhomogeneous corona (right panel).
All of the eigenvalues are positive.}
\end{figure}

It is interesting to compare our model parameters with those used by HKC,
who computed the time-averaged spectra of Cyg~X-1 for a variety of electron density
profiles, similar to the homogeneous and inhomogeneous cloud configurations
studied here. They employed a scattering cloud with a homogeneous central region,
coupled with either a homogeneous or inhomogeneous outer region. The HKC
cloud has a scattering optical thickness $\tau_*=1$ and an electron temperature of
$k T_e = 100\,$keV, whereas we obtain $\tau_* \sim 2-3$ and $k T_e \sim 60\,$keV
(see Table~1). The differences between our model parameters and theirs could be
due to the fact that the observational data analyzed here corresponds to the low/hard
state of Cyg~X-1, whereas HKC compared their model with spectral data from Ling
et al. (1997), acquired while Cyg~X-1 was in its high/soft state, when the source
is known to have a lower optical depth (e.g., Frontera et al. 2001; Malzac 2012;
Del Santo et al. 2013). Furthermore, the values of $\tau_*$ and $T_e$ that we obtain
are very close to those found by Malzac et al. (2008), who also considered the
low/hard state of Cyg~X-1.

\begin{deluxetable}{clccccccrc}
\tabletypesize{\scriptsize}
\tablecaption{Input Model Parameters \label{tbl-1}}
\tablewidth{0pt}
\tablehead{
\colhead{Source}
& \colhead{Model}
& \colhead{$\eta$}
& \colhead{$\Theta$}
& \colhead{$k T_e\,$(keV)}
& \colhead{$\epsilon_{\rm abs}\,$(keV)}
& \colhead{$z_{\rm in}$}
& \colhead{$z_0$}
& \colhead{$t_*\,$(s)}
& \colhead{$\tau_*$}
}
\startdata
Cyg~X-1
&Homogeneous
&2.50
&0.120
&61.3
&1.60
&0.00
&1.00
&0.040
&2.50
\\
Cyg~X-1
&Inhomogeneous
&1.40
&0.122
&62.4
&1.60
&0.12
&0.91
&0.065
&2.97
\\
GX~339-04
&Homogeneous
&4.00
&0.064
&32.7
&0.01
&0.00
&0.78
&0.038
&4.00
\\
GX~339-04
&Inhomogeneous
&2.20
&0.064
&32.7
&0.01
&0.10
&0.60
&0.090
&5.07
\\
\enddata
\end{deluxetable}
\begin{deluxetable}{clcccccccc}
\tabletypesize{\scriptsize}
\tablecaption{Auxiliary Model Parameters \label{tbl-2}}
\tablewidth{0pt}
\tablehead{
\colhead{Source}
& \colhead{Model}
& \colhead{$\dot N_0\,({\rm s}^{-1})$}
& \colhead{$L_{\rm inj}\,({\rm ergs\,s}^{-1})$}
& \colhead{$L_{\rm X}\,({\rm ergs\,s}^{-1})$}
& \colhead{$y_{\rm eff}$}
& \colhead{$\tau_{\rm eff}$}
& \colhead{$\lambda_0$}
& \colhead{$R\,$(cm)}
& \colhead{$D\,$(kpc)}
}
\startdata
Cyg~X-1
&Homogeneous
&$2.00 \times 10^{46}$
&$3.20 \times 10^{36}$
&$2.20 \times 10^{37}$
&1.20
&1.58
&1.20
&$3.00 \times 10^{9}$
&$2.4$
\\
Cyg~X-1
&Inhomogeneous
&$2.70 \times 10^{46}$
&$4.33 \times 10^{36}$
&$2.20 \times 10^{37}$
&1.17
&1.55
&1.25
&$2.73 \times 10^{9}$
&$2.4$
\\
GX~339-04
&Homogeneous
&$5.75 \times 10^{46}$
&$9.21 \times 10^{36}$
&$6.28 \times 10^{37}$
&1.48
&2.40
&0.52
&$4.56 \times 10^{9}$
&$8.0$
\\
GX~339-04
&Inhomogeneous
&$7.00 \times 10^{46}$
&$1.12 \times 10^{37}$
&$6.28 \times 10^{37}$
&1.51
&2.43
&0.51
&$5.94 \times 10^{9}$
&$8.0$
\\
\enddata
\end{deluxetable}

In Table~2 we compare the energy injection rate for the seed photons in our
model, $L_{\rm inj}$, with the time-averaged X-ray luminosity, $L_{\rm X}$, observed
in the low/hard state for the two sources studied here, Cyg~X-1 and GX~339-04.
The injection luminosity is computed using $L_{\rm inj} = \epsilon_0 \, \dot N_0$,
where $\dot N_0$ is the photon injection rate and the seed photon energy is
$\epsilon_0=0.1\,$keV. The values for $L_{\rm X}$ were taken from Cadolle Bel
et al. (2006) for Cyg~X-1, and from Cadolle Bel et al. (2011) for GX~339-04.
We see that the injection luminosity is $\sim 10\%$ of the observed X-ray
luminosity, which is consistent with the values we have obtained for the
effective Compton $y$-parameter.

\subsubsection{Approximate Power-Law X-ray Spectra}

The X-ray spectra plotted in Figures~1 and 2 have a power-law form that extends up to the exponential cutoff, where electron recoil losses become significant. This suggests the existence of an approximate, asymptotic power-law solution, valid in the domain $x \lesssim 1$ (Rybicki \& Lightman 1979). Figures~1 and 2 also include a convergence study, where we compare the results obtained for the steady-state spectra, $\sgreen$, using only the first ($n=0$) term in the series with the fully-converged result obtained using the first 7 terms in the series. The results are essentially indistinguishable, which establishes that the convergence of the series
is extremely rapid. The power-law shape observed for $x \lesssim 1$, combined with the rapid convergence, suggest that we can derive an asymptotic power-law solution by analyzing the first term in the expansion for the observed flux. By analogy with previous work on thermal Comptonization, we expect that the properties of the approximate analytical solution will shed light on the relationship between the first eigenvalue, $\lambda_0$, which determines the spectral slope, and the effective Compton $y$-parameter for the model. We derive the approximate asymptotic power-law solution below, for both the homogeneous and inhomogeneous cloud configurations.

We are interested in photon energies well above the injection energy,
$\epsilon_0=0.1\,$keV, and therefore it follows that $x > x_0$. In this case,
we can combine Equations~(\ref{eq46}) and (\ref{eq64}) to express the
time-averaged X-ray spectrum in the homogeneous corona as
\begin{equation}
\sgreen(x,x_0,z) = {9 \dot N_0 e^{(x_0-x)/2} (x x_0)^{-2} \over 4 \pi R^2 c \,
\Theta^4 (m_e c^2)^3}
\sum_{n=0}^\infty {\Gamma(\sigma-3/2) \sin(\eta\sqrt{\lambda_n})
\over \lambda_n \Gamma(1+2\sigma)\mathscr{I}_n}
\, Y_n(z) M_{2,\sigma}(x_0)W_{2,\sigma}(x)
\label{eq64b}
\ .
\end{equation}
The corresponding result obtained by combining Equations~(\ref{eq85}) and
(\ref{eq99}) in the inhomogenous case is
\begin{equation}
\sgreen(x,x_0,z) = {\dot N_0 e^{(x_0-x)/2} (x x_0)^{-2} \over 2 \pi R^2 \eta c \,
\Theta^4 (m_e c^2)^3}
\sum_{n=0}^\infty {\Gamma(\sigma-3/2) \mathscr{L}_n
\over \mathscr{J}_n \Gamma(1+2\sigma)(1-z_{\rm in}^2)}
\, y_n(z) M_{2,\sigma}(x_0)W_{2,\sigma}(x)
\label{eq64f}
\ .
\end{equation}
Based on Figure~1, we observe that the domain of the power-law shape is
$x_0 < x \lesssim 1$. This suggests that we can employ Equations~(13.1.32),
(13.1.33), (13.5.5), and (13.5.6) from Abramowitz \& Stegun (1970) to implement
the small-argument asymptotic form for the Whittaker functions $M$ and $W$.

We will only evaluate the $n=0$ term in the sum, since it represents a converged
result, according to the results plotted in Figure~1. After some algebra, the
approximate solution obtained in the homogeneous case is
\begin{equation}
\sgreen(x,x_0,z) \approx {9 \dot N_0 \, x_0^{\sigma_0-3/2} \over 8 \pi R^2 c \,
\Theta^4 (m_e c^2)^3}
{\sin(\eta\sqrt{\lambda_0})
\over \lambda_0 \sigma_0 \mathscr{I}_0}
\, {\sin(\eta z\sqrt{\lambda_0}) \over \eta z} \, x^{-\sigma_0-3/2}
\label{eq64c}
\ ,
\end{equation}
where (see Equation~(\ref{eq48}))
\begin{equation}
\sigma_0 \equiv \sqrt{\dfrac{9}{4}+\dfrac{\lambda_0}{3\Theta}}
\ .
\label{eq64index}
\end{equation}
Likewise, in the inhomogeneous case, we obtain
\begin{equation}
\sgreen(x,x_0,z) \approx {\dot N_0 \, x_0^{\sigma_0-3/2} \over 4 \pi R^2 \eta c \,
\Theta^4 (m_e c^2)^3}
{\mathscr{L}_0 \, y_0(z)
\over \mathscr{J}_0 \sigma_0 (1-z_{\rm in}^2)}
\, x^{-\sigma_0-3/2}
\label{eq64d}
\ .
\end{equation}
By substituting either Equation~(\ref{eq64c}) or (\ref{eq64d}) into Equation~(\ref{eq129}),
and setting $z=1$, we can compute the corresponding approximate X-ray spectrum, ${\cal F}_\epsilon(\epsilon)$,
observed at the detector. These results are plotted and compared with the exact solutions
in Figure~4, and it is clear that the power-law approximation is extremely accurate
below the exponential cutoff energy, as expected.

We can obtain further insight into the physical significance of our approximate
power-law solutions by comparing our work with previous results. First, we note
that within the regime of interest here, $x \lesssim 1$, and therefore electron recoil
losses are negligible. This suggests that we can define an effective $y$-parameter by
comparing our work with the corresponding analytical solutions that neglect recoil
losses. This situation was treated by Rybicki \& Lightman (1979),
who obtained power-law solutions to the Kompaneets equation by utilizing an
escape-probability formalism for the spatial photon transport, as an alternative
to the spatial diffusion operator employed here. In our solutions, given by
Equations~(\ref{eq64c}) and (\ref{eq64d}), the power-law index is equal to
$-\sigma_0-3/2$. Setting our result equal to the index $m$ given by Equation~(7.76)
from Rybicki \& Lightman (1979) yields
\begin{equation}
- \sigma_0 - \frac{3}{2} = - \frac{3}{2} -\sqrt{\frac{9}{4} + \dfrac{4}{y_{\rm eff}}}
\label{eq64indexB}
\ ,
\end{equation}
where $y_{\rm eff}$ is the effective Compton $y$-parameter and $\Theta$ is the
dimensionless temperature ratio. Using Equation~(\ref{eq64index}) to substitute
for $\sigma_0$ and solving for $y_{\rm eff}$, we find that
\begin{equation}
y_{\rm eff} = \frac{12 \, \Theta}{\lambda_0}
\ .
\label{eqyeff}
\end{equation}
The values obtained for $y_{\rm eff}$ and $\lambda_0$ in our calculations of the
time-averaged X-ray spectra resulting from distributed (density-weighted) seed photon
injection are reported in Table~2. We generally find that $y_{\rm eff} \sim 1$,
corresponding to unsaturated Comptonization, which is consistent with the
power-law spectra plotted in Figures~1 and 2 (e.g., Sunyaev \& Titarchuk 1980).

It is also interesting to relate the first eigenvalue, $\lambda_0$, to the effective
optical depth, $\tau_{\rm eff}$, traversed by the photons as they propagate through
the scattering corona, and ultimately escape. Referring to the simplified escape-probability
model analyzed by Rybicki \& Lightman (1979), we can apply their Equation (7.41) to write,
in the optically thick case,
\begin{equation}
y = 4 \, \Theta \, \tau_{\rm eff}^2
\label{eqyNR}
\ .
\end{equation}
Setting $y=y_{\rm eff}$ and combining Equations~(\ref{eqyeff}) and (\ref{eqyNR}), we
find that $\tau_{\rm eff}$ and $\lambda_0$ are related via
\begin{equation}
\tau_{\textrm{eff}}=\sqrt{\dfrac{3}{\lambda_0}}
\label{eqtaueff}
\ .
\end{equation}
The results obtained for $\tau_{\rm eff}$ are listed in Table~2. Comparing the values of
$\tau_{\rm eff}$ with the values for $\tau_*$ in Table~1, we conclude that $\tau_{\rm eff} \sim 0.5 \, \tau_*$,
which reflects the fact that the seed photon injection is density weighted, rather than being
localized at the center of the cloud. Hence, on average, photons traverse less optical depth
than is given by $\tau_*$, which is measured from the cloud center.

\begin{figure}[h!]
\captionsetup[subfigure]{labelformat=empty}
\subfloat[]{\includegraphics[scale=0.85]{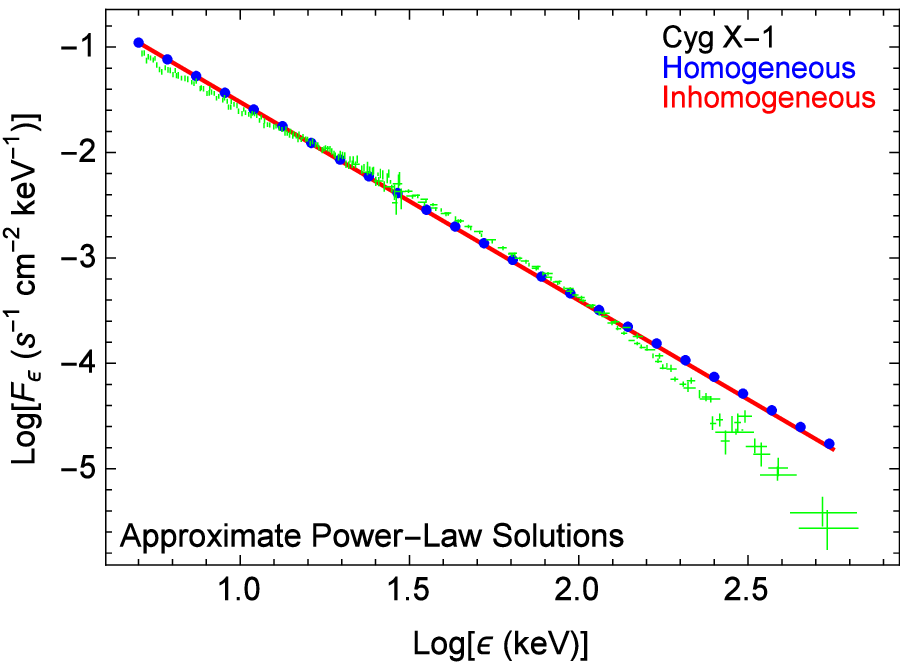}}
\subfloat[]{\includegraphics[scale=0.85]{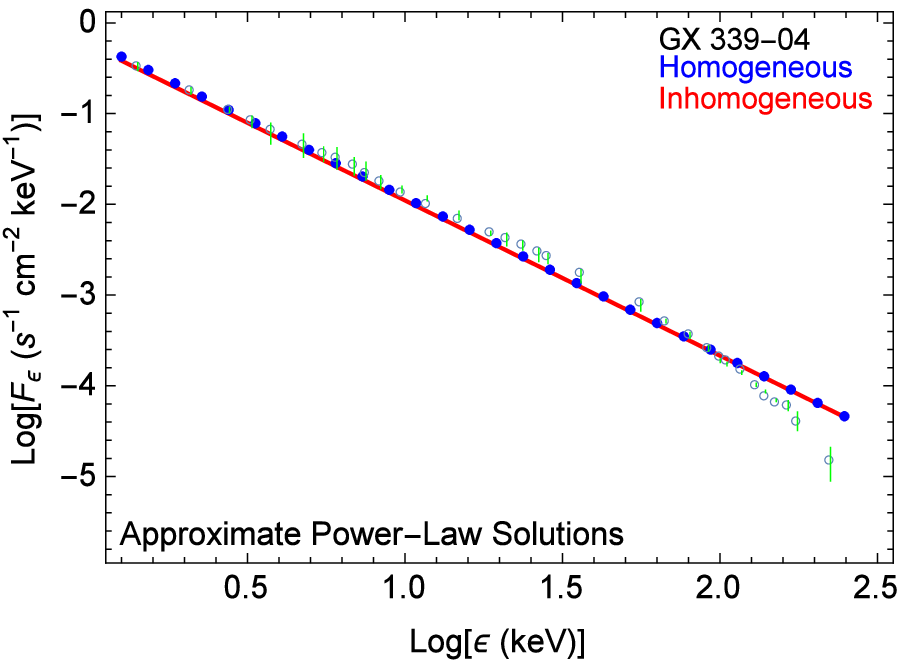}}
\caption{Approximate power-law X-ray spectra, ${\cal F}_\epsilon(\epsilon)$, computed using Equation~(\ref{eq129})
combined with Equation~(\ref{eq64c}) for the homogeneous corona (blue filled circles)
or Equation~(\ref{eq64d}) for the inhomogeneous corona (red solid lines). The results are
compared with the observational data for Cyg~X-1 (left panel) and GX~339-04
(right panel). See the discussion in the text.}
\end{figure}

\subsection{Comparison with Time Lag Data}

In the time-dependent case, the time lags are computed using the Fourier transforms
evaluated at the surface of the cloud, after an impulsive localized transient injects
seed photons with a specified spectrum at a specific radius. This represents a sudden,
low-luminosity flash of radiation that subsequently scatters and Comptonizes throughout
the cloud before the final signal escapes to the observer.

The theoretical prediction for the time lag observed between hard channel energy
$\epsilon_{\rm hard}$ and soft channel energy $\epsilon_{\rm soft}$ at Fourier
frequency $\nu_f$ is computed using the van der Klis et al. (1987) formula
(cf. Equation~(\ref{eq4})),
\begin{equation}
\delta t = \frac{{\rm arg}[S^*(x_{\rm soft},\tilde\omega)
H(x_{\rm hard},\tilde\omega)]}{2\pi \nu_f}
\label{eq130}
\ ,
\end{equation}
where the dimensionless energies $x_{\rm soft}$ and $x_{\rm hard}$ are defined by
\begin{equation}
x_{\rm soft} \equiv \frac{\epsilon_{\rm soft}}{k T_e} \ , \qquad
x_{\rm hard} \equiv \frac{\epsilon_{\rm hard}}{k T_e}
\ .
\label{eq130b}
\end{equation}
The Fourier transforms of the soft and hard channel time series are computed
using
\begin{equation}
S(x_{\rm soft},\tilde\omega) = F(x_{\rm soft},\tilde\omega) \ , \qquad
H(x_{\rm hard},\tilde\omega) = F(x_{\rm hard},\tilde\omega)
\ ,
\label{eq130c}
\end{equation}
where $F$ represents the Fourier transform radiated at the surface
of the coronal cloud, at radius $r=R$ ($z=1$). We assume that the observed
time lags are the result of the time-dependent Comptonization of seed
photons injected with either a monochromatic or bremsstrahlung initial
energy distribution. Our results for the homogeneous and inhomogeneous
Fourier transforms in the case of monochromatic photon injection are given
by Equations~(\ref{eq79}) and (\ref{eq119}), respectively, and our
results for the homogeneous and inhomogeneous Fourier transforms
in the case of bremsstrahlung injection are both covered by Equation~(\ref{eq124}).
In the case of bremsstrahlung injection, we must also impose a low-energy
self-absorption cutoff at energy $\epsilon = \epsilon_{\rm abs}$ in order to
avoid producing an infinite number of seed photons.

All of our analytical formulas for the Fourier transform are expressed in terms
of the dimensionless Fourier frequency, $\tilde\omega$, which is related to
the dimensional Fourier frequency, $\nu_f$, measured in Hz, via (see
Equation~(\ref{eq35}))
\begin{equation}
\tilde\omega = 2 \pi \nu_f t_*
\label{eq131}
\ ,
\end{equation}
where the scattering time, $t_*=\ell_*/c$, is equal to the mean-free time at the
outer edge of the cloud. The value of $t_*$ is related to the cloud radius $R$
and the value of $\eta$ via (see Equation~(\ref{eq20}))
\begin{equation}
t_* = \frac{\ell_*}{c} = \frac{R}{\eta \, c}
\label{eq131b}
\ .
\end{equation}

Once the values for the temperature parameter $\Theta$, the scattering parameter
$\eta$, and the inner radius $z_{\rm in}$ have been tied down via comparison of
the observed time-averaged spectrum with the theoretical steady-state spectrum for
a given source, the next step is to vary the values of the cloud radius, $R$, and
the bremsstrahlung self-absorption energy, $\epsilon_{\rm abs}$, until we achieve
reasonable qualitative agreement between the theoretical time lags and the observed
time lags. This allows us to translate between the dimensionless Fourier frequency
$\tilde\omega$ and the dimensional frequency $\nu_f$ using Equation~(\ref{eq131}),
with the scattering time $t_*$ computed using Equation~(\ref{eq131b}). We consider
several different scenarios for the calculation of the X-ray time lags below and
compare the results with the observational data for Cyg~X-1 and GX~339-04.

\subsubsection{Monochromatic Injection in Inhomogeneous Corona}

When the injected spectrum is monochromatic, or nearly so, and the
injection takes place in a homogeneous cloud, all of the authors
who have examined the problem agree that the resulting time lags are
independent of Fourier frequency, in contradiction to the observations
(e.g. Miyamoto 1988, HKC, KB). Hence it is interesting to explore the
consequences of altering the cloud configuration in our model to treat
monochromatic seed photon injection in an {\it inhomogeneous} corona,
with electron number density distribution $n_e(r) \propto r^{-1}$, which was
also considered by HKC. Since the injected seed photons are monochromatic,
with energy $\epsilon_0=0.1\,$keV, we must use the Fourier transform
Green's function, $\Green$, to compute the time lags by combining
Equations~(\ref{eq119}), (\ref{eq130}), and (\ref{eq130c}). The time lags
resulting from monochromatic injection in an inhomogeneous cloud are
plotted as a function of the Fourier frequency $\nu_f$ and compared with
the Cyg~X-1 data from Nowak et al. (1999) in Figure~5 for both large and
small cloud radii. The channel energy values used are $\epsilon_{\rm soft}=2\,$keV
and $\epsilon_{\rm hard }=11\,$keV, which correspond to the channel-center energies
used in the analysis of the observational data. It is clear that the model results
do not fit the data very well for either value of the cloud radius. Note that the
shape of the time lag curves exhibits the same trend as the data, but the
magnitude is too large. This is a result of the long upscattering time required
for the soft disk seed photons to reach the soft and hard channel energies.

HKC also computed time lags for monochromatic injection in an
inhomogeneous cloud, but they were able to fit the observational data,
in contrast to our results. However, in order to qualitatively match the observed
time lags, HKC had to adopt an outer cloud radius of $\sim1\,$light-second
($3 \times 10^{10}\,$cm), which is an order of magnitude larger than the cloud
radii implied by our model. The discrepancy between the model results may be
due to the fact that their cloud is optically thin, whereas our cloud is optically
thick. The values for the optical depth derived here are consistent with those
obtained during the low/hard state of Cyg~X-1 by Malzac et al. (2008),
Malzac (2012), Del Santo et al. (2013), and Frontera et al. (2001). Unfortunately,
we can't use our model to explore the region of parameter space studied by HKC
because the corona must be optically thick in order to justify the diffusion
approximation employed in our approach.

\begin{figure}[h!]
\vspace{0.0cm}
\centering
\includegraphics[height=8cm]{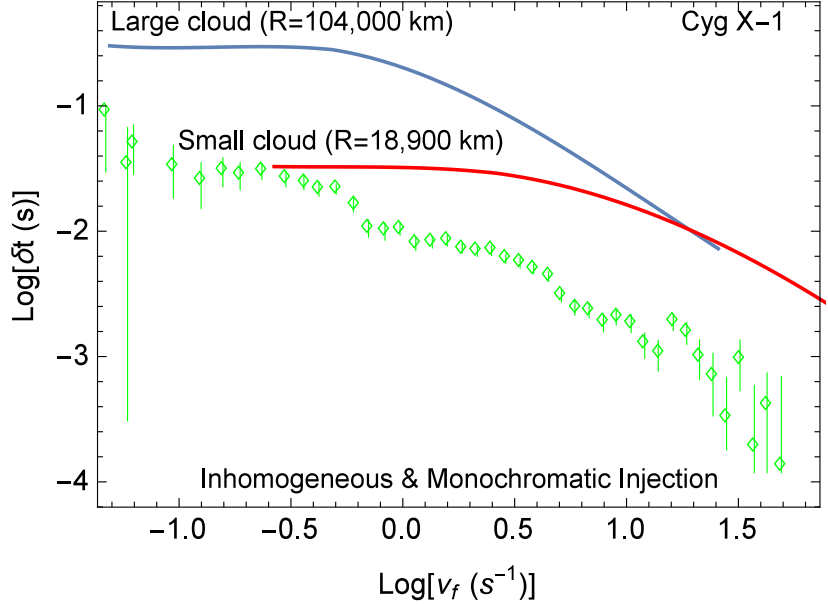}
\caption{Theoretical time lag profiles resulting from monochromatic injection in an
inhomogeneous cloud, with electron number density profile $n_e(r) \propto r^{-1}$,
compared with the Cyg~X-1 time lag data from Nowak et al. (1999). The source 
was in the low/hard state during the observation. The time lags
are computed by combining Equations~(\ref{eq119}), (\ref{eq130}), and (\ref{eq130c}),
and the channel energies used in the theoretical calculations are $\epsilon_{\rm soft}
= 2\,$keV and $\epsilon_{\rm hard } = 11\,$keV. The photon injection energy is
$\epsilon_0=0.1\,$keV.}
\end{figure}

\subsubsection{Variation of Seed Photon Distribution}

It is apparent from Figure~5 that monochromatic injection into an inhomogeneous
corona is unable to generate good agreement with the time lag data. Furthermore, it
has been previously established by Miyamoto (1988), HKC, and KB that monochromatic
injection into a homogeneous cloud also fails to agree with the data. Hence, it
is interesting to use our new formalism to explore the alternative hypothesis of
broadband (bremsstrahlung) seed photon injection, rather than monochromatic injection.

The bremsstrahlung-injection time lags are computed by combining Equations~(\ref{eq124}),
(\ref{eq130}), and (\ref{eq130c}), and the model parameters are varied until reasonable
qualitative agreement with the observational data is achieved. We plot the theoretical
bremsstrahlung-injection time lags as a function of the Fourier frequency $\nu_f$ in Figure~6,
using both the homogeneous and inhomogeneous coronal cloud models. The results are
compared with the observational data for Cyg~X-1 and GX~339-04 taken from Nowak et al.
(1999) and Cassatella et al. (2012), respectively. The corresponding physical parameters
are listed in Table 1, and the channel energies used in the theoretical calculations are
$\epsilon_{\rm soft}=2\,$keV and $\epsilon_{\rm hard}=11\,$keV for Cyg X-1, and
$\epsilon_{\rm soft}=2\,$keV and $\epsilon_{\rm hard}=10\,$keV for GX~339-04, which
correspond to the channel-center energies used in the observational
calculations of the time lags. The low-energy self-absorption cutoff is set at
$\epsilon_{\rm abs}=1.6\,$keV for Cyg~X-1 and at $\epsilon_{\rm abs}=0.01\,$keV
for GX~339-04. In the case of the homogeneous corona, the eigenvalues
$\lambda_n$ for the Fourier transform solution are the same real values
obtained in the analysis of the time-averaged (quiescent) spectrum, which are plotted in the
left-hand panel in Figure~3. In the case of the inhomogeneous corona,
the eigenvalues $\lambda_n$ are complex, and are plotted in Figure~7.

We find that in order to match the observational time lag data, the impulsive injection
of the bremsstrahlung photons must occur near the outer edge of the cloud, with
$z_0 \lesssim 1$. The transient that produces the soft seed photons is not treated in
detail here, but we note that the outer edge of the corona is a region which the disk
suddenly expands in the vertical direction, possibly leading to various types of plasma
instabilities. In particular, the abrupt change in magnetic topology may generate rapid
reconnection events that can result in the injection of a significant population of soft
seed photons via bremsstrahlung emission (e.g., Poutanen \& Fabian 1999).

In contrast with the behavior of the monochromatic injection scenario studied
by Miyamoto (1988), the results depicted in Figure~6 show that in the case
of broadband (bremsstrahlung) seed photon injection into either a
homogeneous or inhomogeneous cloud, Comptonization can produce Fourier
frequency-dependent time lags that agree with the observational data for both
Cyg~X-1 and GX~339-04. The diminishing time lags at high Fourier frequency
are explained as a natural results of the prompt escape of broadband seed
photons, combined with the delayed escape of upscattered Comptonized
photons over longer timescales.

\begin{figure}[h!]
\captionsetup[subfigure]{labelformat=empty}
\subfloat[]{\includegraphics[scale=0.85]{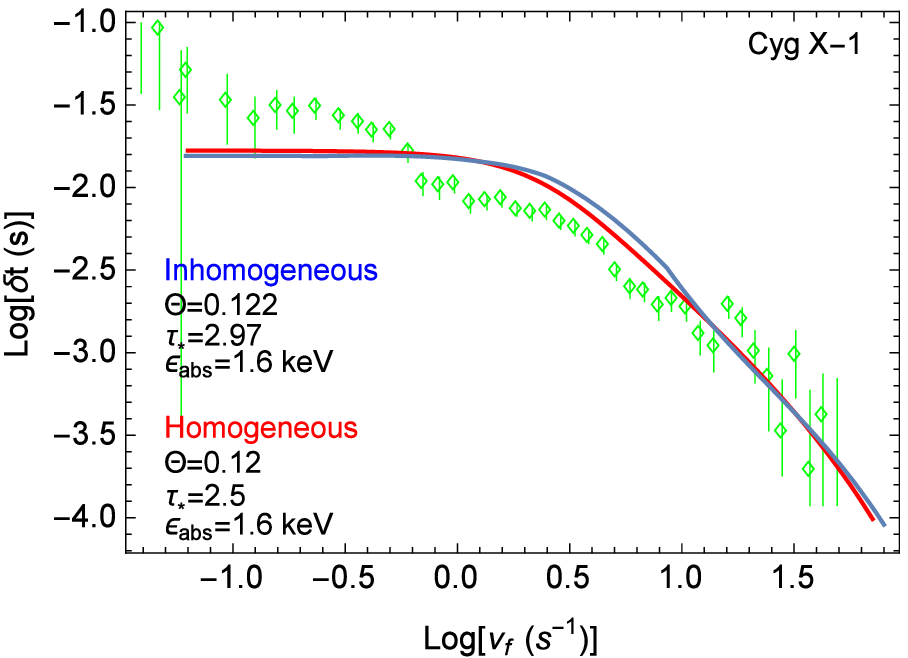}}
\subfloat[]{\includegraphics[scale=0.85]{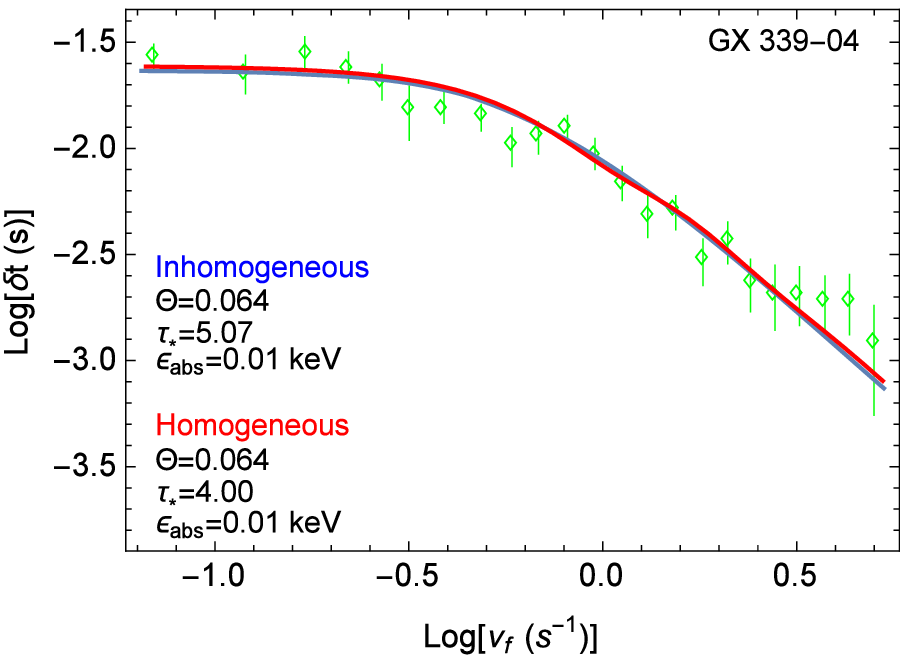}}
\caption{Theoretical time lag profiles for bremsstrahlung seed photon
injection in a homogeneous corona (red) and an inhomogeneous corona
(blue), compared with the data for Cyg~X-1 (left panel) from Nowak et al.
(1999), and the data for GX~339-04 (right panel) from Cassatella et al. (2012).
Each source was observed in the low/hard state. See Section~6.2.3 and 
Figure~8 for a discussion of the convergence properties.}
\end{figure}

\begin{figure}[h!]
\captionsetup[subfigure]{labelformat=empty}
\subfloat[]{\includegraphics[scale=0.85]{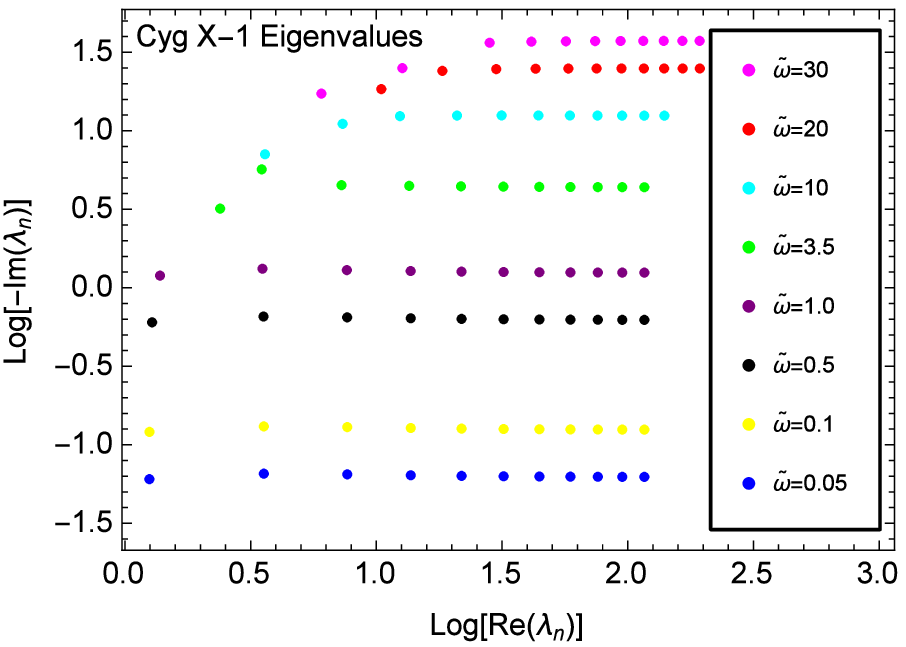}}
\subfloat[]{\includegraphics[scale=0.85]{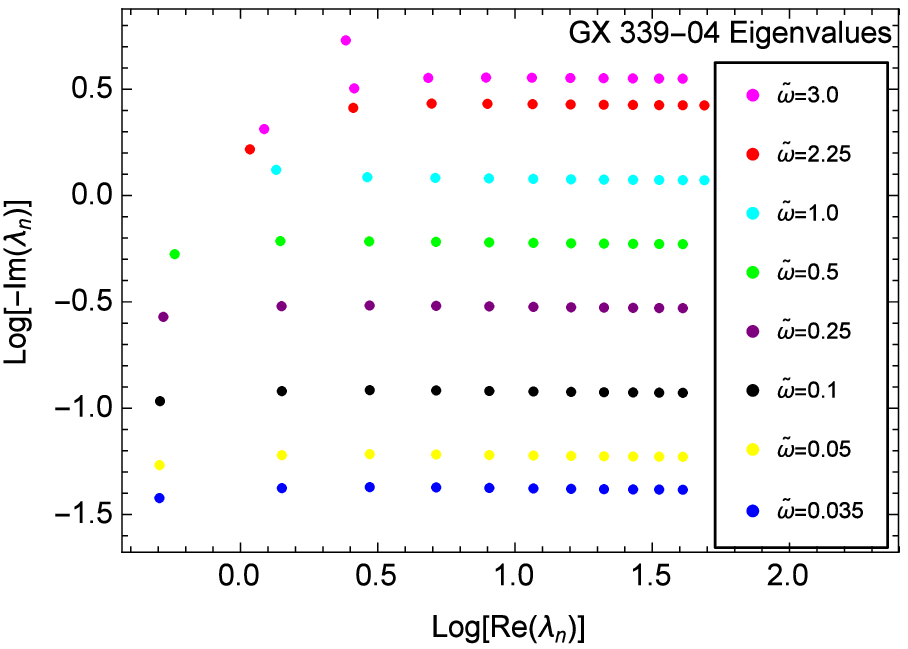}}
\caption{Complex eigenvalues, $\lambda_n$, for the Fourier transform
in the inhomogeneous case. Left panel is for Cyg~X-1 and right panel
is for GX~339-04. Note that the imaginary part of $\lambda_n$ is always
negative, and therefore we change the sign before taking the log. The colors
refer to the indicated values of the dimensionless Fourier frequency
$\tilde\omega$, and the sequences running from left to right represent
the values of $\lambda_0$ through $\lambda_{10}$.}
\end{figure}

This indicates that the critical quantities for determining the shape of the
time-lag profile are the overall optical thickness of the cloud and its temperature,
which have nearly the same values in the homogeneous and inhomogeneous
corona models. We therefore conclude that the actual configuration of the cloud
(i.e. the detailed radial variation of the electron number density) is not well
constrained by either the observations of the time lags or the observations
of the time-averaged X-ray spectrum, and indeed, either cloud configuration works
equally well, although there is a slight difference in the resulting cloud radius $R$,
as indicated in Table~2.

\subsubsection{Convergence of Time Lags}

In our model, the time lags are computed based on analytical expressions
for the Fourier transform of the emitted radiation spectrum. Since these expressions
are stated in terms of series expansions, it is important to examine the convergence
of the results obtained for the time lags as one increases the truncation level of the
series. Obviously, rapid smooth convergence is desirable.

In Figure~8, we present a convergence study of the theoretical time lags computed
using the models for Cyg~X-1 and GX~339-04, based on both the homogeneous and
inhomogeneous cloud configurations. In each panel, the black curves represent the time
lags evaluated using only the first term in the expansions, and the red and blue curves
represent fully converged results, where no significant change will occur upon the addition
of another term. The red and blue curves are the same as the final results for the time lags
plotted in Figure~6. The time lags generally require about 20 terms to fully converge,
whereas the expansions for the time-averaged spectra converge immediately (see Figures~1
and 2).

\begin{figure}[h!]
\vspace{0.0cm}
\centering
\includegraphics[height=11cm]{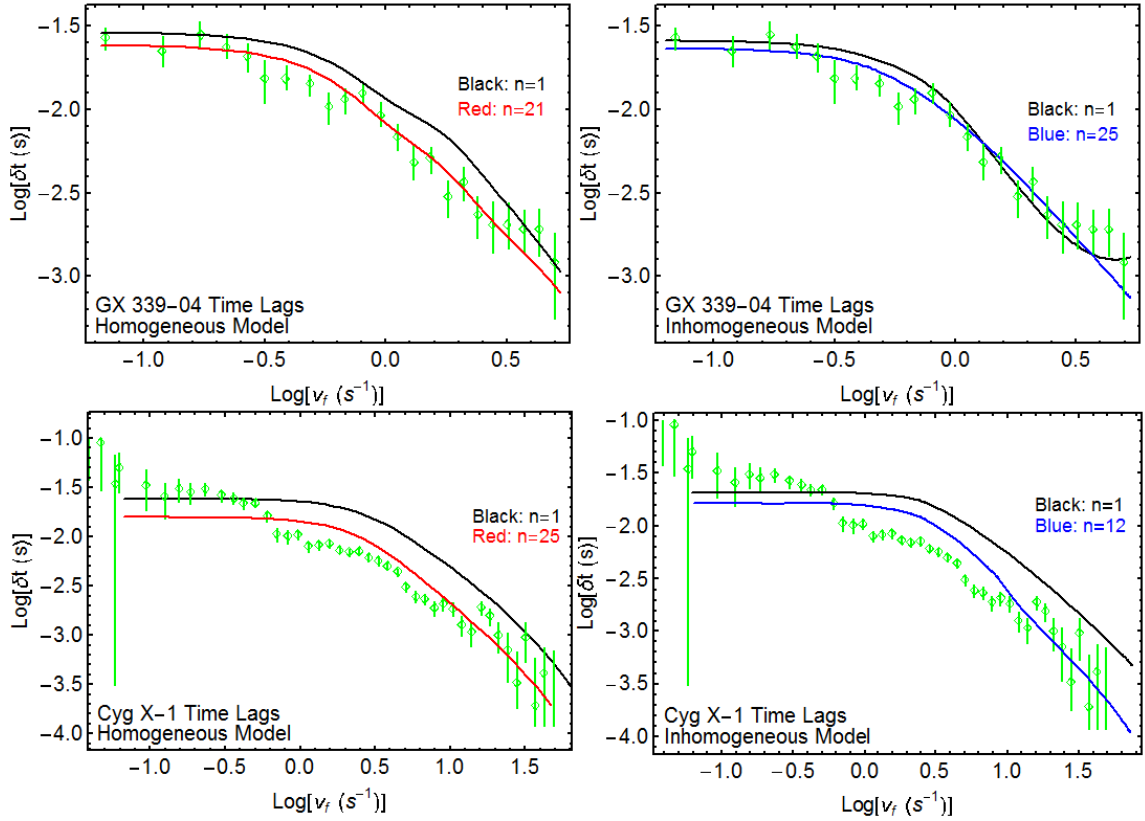}
\caption{Convergence study of the theoretical time lags for Cyg~X-1 and
GX~339-04 computed using either the homogeneous or the inhomogeneous
cloud model.
The number of terms used in the series expansions for the Fourier transforms
is indicated for each curve. The red and blue curves correspond to the final results
plotted in Figure~6.}
\end{figure}

\section{Discussion and Conclusion}

We have obtained the exact analytical solution for the problem of
time-dependent thermal Comptonization in a spherical scattering corona,
based on two different electron density profiles. By working in the
Fourier domain, we have obtained a closed-form expression for the
Green's function corresponding to the injection of monochromatic seed
photons into a cloud at a single radius and time. The radiated Fourier
transform, evaluated at the surface of the cloud, can be directly
substituted into the time lag formula introduced by van der Klis et al.
(1987) in order to compute the predicted dependence of the lags on the
Fourier frequency for any selected X-ray channel energies. In our
approach, the time-averaged X-ray spectrum and the time lags are both
computed using the same set of physical parameters to describe the
properties of the scattering cloud, and therefore our formalism
represents an integrated model that fully describes the high-energy
spectral and timing properties of the source.

\subsection{Relation to Previous Work}

The study presented by HKC is similar to ours, although their methodology and
input assumptions are somewhat different. HKC focused exclusively on a single
injection scenario, namely the injection of essentially monoenergetic, low-temperature
blackbody seed photons at the center of the scattering cloud. Based on this injection
spectrum, they concluded that the observed time lag behavior in Cyg~X-1 could
not be reproduced unless the electron number density profile was inhomogeneous,
with $n_e(r) \propto r^{-1}$ for example. In this case, although the predicted
time lags fit the observed dependence on the Fourier period, the resulting
dimensions of the cloud are so large that the requisite heating is difficult to
accomplish based on any of the standard dissipation models.

Another notable difference between the work of HKC and the results developed
here is that we have obtained a set of exact mathematical solutions,
whereas HKC utilized a numerical Monte Carlo simulation method. This distinction is
important, because by exploiting the exact solution for the Fourier
transform of the Green's function, we are able to explore a much
wider range of injection scenarios, in which we can vary both the location of the
initial flash of seed photons, and its spectral distribution. Based on our analytical
formalism, we are able to confirm the results of HKC regarding monochromatic
injection, but we have also generalized those results by exploring
the implications of varying the seed photon injection radius and
spectrum. We find that the injection of {\it broadband} (bremsstrahlung)
seed photons relatively close to the surface of a homogeneous or
inhomogeneous cloud can fit the observed time lag profiles at least as
well as the HKC model does, but with a cloud size an order of magnitude
smaller. In Section~7.2 we discuss the physical reasons underlying
the success of the bremsstrahlung injection scenario.

The treatment of electron scattering in our work differs from that utilized by HKC,
since we have adopted the Thomson cross section,
whereas HKC implemented the full expression for the Klein-Nishina cross section.
In principle, utilization of the Klein-Nishina cross section would be expected to
affect the hard time lags, due to the quantum reduction in the scattering probability at
high energies. However, for the photon energy range of interest here, $\sim 0.1-10\,$keV,
combined with our maximum electron temperature, $kT_e=62.4\,$keV, not many photons
are likely to sample the reduced cross section, which requires an incident photon
energy exceeding $500\,$keV as seen in the rest frame of the electron.
Hence it seems surprising that HKC observed a significant change in
the normalization of their computed time lags when they
adopted the Klein-Nishina cross section instead of the Thomson value. We suspect that
this may be due to the somewhat
higher electron temperature they used, $kT_e=100\,$keV.

To explore this question
quantitatively, we can compute the fraction of electrons such that an incident photon
of a given energy in the lab frame exceeds $500\,$keV in the electron's rest frame.
The relevant thermal distribution function for the calculation is the relativistic
Maxwell-J\"uttner distribution, given by (e.g., Ter Haar \& Wergeland 1971; Hua 1997)
\begin{equation}
f_{\rm MJ}(\gamma) \equiv \frac{\gamma\sqrt{\gamma^2-1}}{\Theta K_2(1/\Theta)}
\exp\left(-\frac{\gamma}{\Theta}\right)
\end{equation}
where $\Theta\equiv kT_e/(m_e c^2)$ and $K_2$ denotes the modified Bessel function of
the second kind. The probability that a randomly-selected electron has a Lorentz factor
in the range between $\gamma$ and $\gamma+d\gamma$ is equal to $f_{\rm MJ}(\gamma) d\gamma$.

In order to compute an upper bound on the probability of generating a scattering in the
Klein-Nishina regime, we shall focus on the most energetic
possible collision scenario, which is a head-on collision between the electron and
the photon. In this case, the incident photon energy in the electron's rest frame, $E'_0$, is given by
\begin{equation}
E'_0 = E_0 \left(\frac{1+\beta}{1-\beta}\right)^{1/2} \ , \qquad \beta^2 = 1 - \frac{1}{\gamma^2} \ ,
\end{equation}
where $E_0$ is the incident photon energy in the lab frame.
By integrating the Maxwell-J\"uttner distribution, we can compute the probability, $P$,
that a randomly-selected electron has sufficient energy to create the required
incident photon energy of at least $500\,$keV in the rest frame. The probability is given by
\begin{equation}
P = \int_{\gamma_0}^\infty f_{\rm MJ}(\gamma) \, d\gamma \ ,
\label{MaxJutProb}
\end{equation}
where the lower bound $\gamma_0$ is the root of the equation
\begin{equation}
500\,{\rm keV} = E_0 \left(2 \gamma_0^2 - 1 + 2 \gamma_0 \sqrt{\gamma_0^2-1} \right)^{1/2} \ .
\end{equation}

Setting the incident photon energy $E_0 = 100\,$keV as an extreme example, we find
that the lower bound is $\gamma_0=2.6$. Adopting the HKC temperature value,
$kT_e=100\,$keV, we obtain $\Theta=0.2$, in which case the probability
given by Equation~(\ref{MaxJutProb}) is
$P=3.1 \times 10^{-3}$. This probability may be large enough to explain the variation of the HKC time lag
results observed when they switched between the Thomson cross section and the Klein-Nishina cross section,
if some of the photons inverse-Compton scatter up to high enough energies to sample the
Klein-Nishina regime, before returning to lower energies via Compton scattering.
We can also compute the scattering probability $P$ based on the maximum electron temperature
that we have adopted in our applications, $kT_e=62.4\,$keV, which yields $\Theta=0.122$.
In this case, one finds that the Maxwell-J\"uttner integration
gives $P=2.6 \times 10^{-5}$, which is much smaller than the HKC result. Hence
we conclude that utilization of the Klein-Nishina cross section would probably not
make a significant difference in our applications. However, we can't reach any definitive conclusions
about this question using the model developed here since it is based on
the assumption of Thomson scattering in the electron's rest frame.

\subsection{Formation of the Light Curves}

The somewhat surprising difference between the time lag profiles
produced when the injection spectrum has a monoenergetic shape versus a
broadband shape can be explored by using the inverse Fourier transform
to compute the time-dependent light curves for the hard and soft energy
channels in the two cases. To accomplish this, we must make use of the
inversion integral (cf. Equation~(\ref{eq34}))
\begin{equation}
f(x,z,p) = \dfrac{1}{2 \pi} \int_{-\infty}^\infty
e^{-i \tilde{\omega} p} F(x,z,\tilde{\omega}) \, d\tilde{\omega}
\label{eq132}
\ ,
\end{equation}
where $F$ is the Fourier transform computed using either the
monochromatic injection Green's function solution
(Equation~(\ref{eq79}) for the homogeneous cloud, or
Equation~(\ref{eq119}) for the inhomogeneous cloud), or the
bremsstrahlung injection solution (the homogeneous and inhomogeneous
cases are both computed using Equation~(\ref{eq124})). Evaluation of
Equation~(\ref{eq132}) requires numerical integration since the
inversion integral cannot be performed analytically. We therefore focus
on a few simple examples in order to illustrate the dependence of the light
curves on the injection model.

In Figure~9, we plot the hard and soft channel light curves computed
using Equation~(\ref{eq132}) for the case of a homogeneous cloud
experiencing impulsive injection of either low-energy monochromatic seed
photons or broadband (bremsstrahlung) seed photons. The parameters
describing the monochromatic injection scenario are temperature
$\Theta=0.12$, injection location $z_0=1$, injection energy
$\epsilon_0=0.1\,$keV, soft channel energy $\epsilon_{\rm soft}=2\,$keV,
and hard channel energy $\epsilon_{\rm hard}=10\,$keV. In the case of
bremsstrahlung injection, we set $\Theta=0.12$, $z_0=1$, $\epsilon_{\rm
abs}=0.1\,$keV, $\epsilon_{\rm soft}=2\,$keV, and $\epsilon_{\rm
soft}=10\,$keV. One can immediately identify the characteristic Fast
Rise Exponential Decay (FRED) shape (e.g., Sunyaev \& Titarchuk 1980)
for each channel signal. As expected, the hard channel curve is delayed
in time relative to the soft channel curve due to upscattering, but the
detailed relationship between the two light curves depends qualitatively
on whether the injection spectrum is monochromatic or broadband.

One clearly observes that the two FRED curves resulting from
monochromatic injection in a homogeneous cloud are of the same shape,
and are simply shifted by a perfect delay with respect to one another
on all timescales (see Figure~9). This yields a constant time lag
across all Fourier frequencies (or periods), in agreement with the Miyamoto
result that HKC and KB have confirmed. Our physical understanding of
this behavior is as follows. Since all of the initial photons start with
the same energy in the monochromatic case, the time lag is purely
a result of Compton reverberation, where the upscattering timescale
is proportional to the logarithm of the ratio of the hard to soft energies
(Payne 1980). Based on this simple example, we conclude that monochromatic
injection anywhere in a homogeneous cloud cannot produce Fourier
frequency-dependent time lags, in contradiction with the observational data.

\begin{figure}[h!]
\vspace{0.0cm}
\centering
\includegraphics[height=6cm]{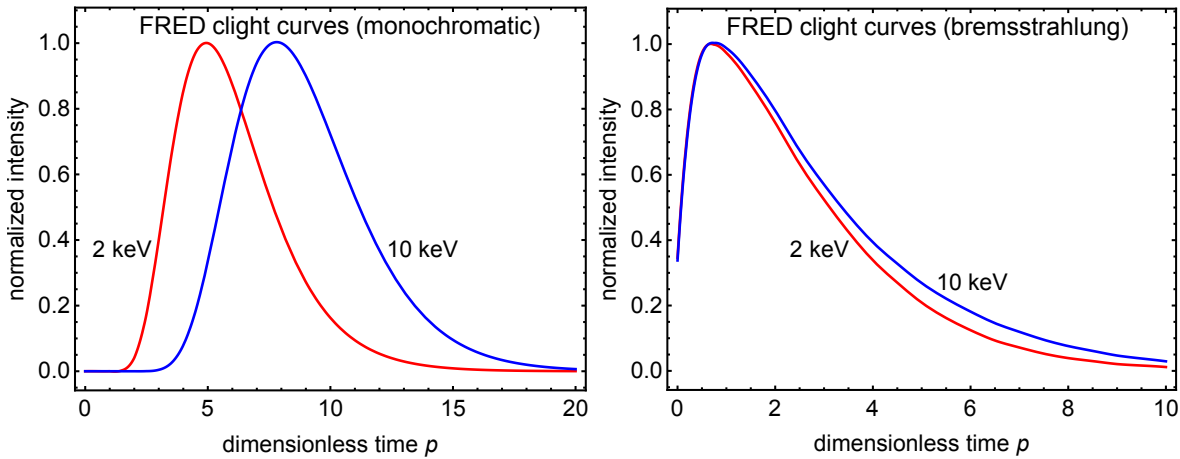}
\caption{FRED curves from monochromatic and bremsstrahlung injection in
a homogeneous cloud. In each case, the red curve represents the soft energy
channel, set at 2~keV, and the blue curve denotes the hard channel, set at 10~keV.
The normalized intensity in each channel shows the relative lag.}
\end{figure}

The relationship between the two FRED light curves plotted in Figure~9 for the
case of bremsstrahlung injection is qualitatively different from the monochromatic
example. In this case, the initial fast rise in both channels is coherent, meaning
that the hard and soft channel signals track each other relatively closely. This
results in a small time lag at high Fourier frequencies, because the fast rise
portion of each curve represents the most rapid variation in the system.
Physically, this part of the process corresponds to the prompt escape of
``pristine'' bremsstrahlung seed photons that are almost unaffected by
scattering. Because bremsstrahlung is a broadband emission mechanism,
both hard and soft photons exist in the initial distribution, and the prompt escape
is therefore coherent across the energy channels. This is, of course, {\it not} true
in the case of low-energy monochromatic injection, because in that scenario, photons
require sufficient time to upscatter into both the soft and hard energy channels.

At longer timescales (smaller Fourier frequencies) in the bremsstrahlung
case, the hard light curve approaches a delayed version of the soft
light curve, reflecting the time it takes for the photons to Compton
upscatter to the hard channel energy. This part of the process is
similar to the monochromatic case, and indeed, we see that the time lags
level off to a plateau at small Fourier frequencies, just as in the
monochromatic example. To summarize, the overall behavior of the
bremsstrahlung injection model matches the observational data much more
closely then does the monochromatic injection scenario because of the
combination of prompt escape (the fast rise part of the light curves)
along with Compton reverberation (exponential decay) on longer
timescales. This explains the origin of the qualitative difference in the
behavior of the time lags at high Fourier frequencies exhibited in the
monochromatic and bremsstrahlung injection scenarios, depicted in
Figures~5 and 6, respectively.

\subsection{Coronal Temperature}

Both our model and that analyzed by HKC require the presence of hot
electrons with temperature $T_e \sim 10^8\,$K at distances $r \sim 10^3\,GM/c^2$
from the black hole. This temperature distribution is consistent with a substantial
number of studies that focus on energy transport in inefficient accretion flows, with
accretion rates that are significantly sub-Eddington, as first established by Nayaran
\& Yi (1995) in the context of the original, self-similar Advection Dominated Accretion
Flow (ADAF) model. Similar results for the temperature distribution were later obtained
using more complex numerical simulations by Oda et al. (2012), Rajesh \&
Mukhopadhyay (2010), Yuan et al. (2006), Mandal \& Chakrabarti (2005), Liu et al. (2002),
R\'o\.za\'nska \& Czerny (2000), and You et al. (2012). In these hot ADAF flows, the
density in the outer region is so low that bremsstrahlung and inverse-Compton
cooling are very inefficient. The lack of efficient cooling drives the electron
temperature in the corona close to the virial value, out to distances of hundreds or
thousands of gravitational radii from the black hole, in agreement with the
temperature profiles assumed here.

In the study presented here, we have assumed that the electron scattering corona is isothermal
in order to accomplish the separation of variables that is required to obtain analytical
solutions to the radiation transport equation. The resulting analytical solutions allow
us to determine the physical properties of the scattering corona in a given source by
computing the time-averaged X-ray spectrum and the time-lag profile and comparing the
theoretical results with the observational data. The assumption of an isothermal corona
is roughly justified by studies indicating that the temperature does not vary by more than
a factor of a few across the corona (e.g., You et al. 2012; Schnittman et al. 2013). Nonetheless,
it is worth asking whether our results would be significantly modified in the presence of a
coronal temperature gradient.

If the electron temperature varied across the corona, then in general one would
expect the plasma to be hotter in the inner region, where the density is likely to be higher
as well. In this scenario, the photons in the hot inner region would Compton upscatter
faster than those in the cooler outer region, but they would spend more time (on average)
scattering through the cloud before escaping due to the greater optical depth in the inner
region. We estimate that these two effects would roughly offset each other, leaving the time
lag profile close to the isothermal result derived here, if the temperature were set equal to
the average value in the corona. Hence we predict that the results obtained for the time
lags in the presence of a temperature gradient would be qualitatively similar to those
obtained here using the isothermal assumption. Moreover, while the electrons may approach the virial
temperature in the outer region, it is likely that in the inner region, the electron temperature is 
thermostatically controlled by Compton scattering (e.g., Sunyaev \& Titarchuk 1980; Shapiro, 
Lightman \& Eardley 1976). The combination of these two effects will tend to produce a
relatively high, but uniform, electron temperature distribution, as we have assumed here.

\subsection{Time Varying Coronal Parameters}

If the transients responsible for producing the observed X-ray time lags in accreting
black hole sources are driven by the deposition of a large amount of energy, then the properties
of the corona (temperature, density) would be expected to respond. If this response occurs
on time scales comparable to the diffusion time for photons to escape from the corona, then
the resulting time lag profiles would be modified compared with the results obtained here,
since we assume that the properties of the corona remain constant. Malzac \& Jourdain
(2000) have considered the possible variation of the coronal properties during X-ray
flares using a non-linear Monte Carlo simulation to study the flare evolution as a function
of time, along with the associated variation of the temperature and optical depth in the
corona. They do not compute Fourier time lags, but they do present simulated light curves
in the soft and hard energy channels. In their model, the flares are driven by a sudden
increase in the disk's internal dissipation, which produces a large quantity of soft photons.
The temperature and optical depth of the corona change self-consistently during the transient,
and then return to the equilibrium state. They find that hard time lags are produced during the
flare if the energy deposition is substantial.

The approach taken by Malzac \& Jourdain (2000) is based on the pulse-avalanche
model of Poutanen \& Fabian (1999). The model does not explicitly include Compton
upscattering as a contributor to the time-lag phenomenon, nor was the significance of
the injection spectrum considered. The simulated light curves generated by Malzac
\& Jourdain (2000) sometimes display temporal dips, but the time dependence doesn't
seem to resemble that observed during the transients in Cyg~X-1. Since these authors
do not compute Fourier time lags, it is difficult to directly compare their results with ours.
However, we note that the transients under study here represent relatively small variations
in the X-ray luminosity, which suggests that the energy deposition may not be large enough
to significantly alter the large-scale properties of the scattering corona during the time it
takes the photons to diffuse out of the cloud (Nowak et a. 1999; Cassatella et al. 2012).
This supports our assumption that the temperature and density of the corona remain
essentially constant during the formation of the observed time lags.

\subsection{Conclusion}

Our goal in this paper is to develop an integrated model, based on
the diffusion and thermal Comptonization of seed photons in an optically-thick
scattering cloud, that can naturally reproduce both the observed X-ray spectra
and the time lags for Cyg~X-1 and GX~339-04 using a single set of cloud
parameters (density, radius, temperature). We have derived and presented
a new set of exact mathematical solutions describing the Comptonization of
seed photons injected into a scattering cloud of finite size that is either
homogeneous, or possesses an electron number density that varies with
radius as $n_e(r) \propto r^{-1}$. The results developed here include new
expressions for (a) the Green's function describing the radiated
time-averaged X-ray flux (corresponding to the reprocessing of continually
injected monochromatic seed photons), (b) the Green's function for the
Fourier transform of the time-dependent radiation spectrum resulting from
the impulsive injection of monochromatic seed photons, and (c) the associated
X-ray Fourier time lags.

By exploiting the linearity of the fundamental transport equation, we used our
results for the Green's function to explore a variety of seed photon injection scenarios.
One of our main conclusions is that the integrated model can successfully explain
the data regardless of the cloud configuration (homogeneous or inhomogeneous),
provided the optical thickness and the temperature are comparable in the two models,
as expected based on the Compton reverberation scenario (Payne 1980). Our results
demonstrate that the bremsstrahlung injection model fits the observational time-lag
data reasonably well for both Cyg~X-1 and GX~339-04, whether the scattering corona
is homogeneous or inhomogeneous. We therefore conclude that the constant time
lags found by HKC in the homogeneous cloud configuration were the result of their
utilization of a quasi-monochromatic (low-temperature blackbody) injection spectrum
for the seed photon distribution.

The injection location in our model is different from that considerd by HKC, who
assumed that the seed photons were always injected at the center of the spherical
cloud. In our model, the injection location is arbitrary, and we find that the best
agreement with the time lag data is obtained when the injection is relatively close
to the surface of the cloud, so that the prompt escape of some of the unprocessed
bremsstrahlung seed photons is able to explain the diminishing time lags
observed at high Fourier frequencies. At longer timescales, the standard
thermal Comptonization process sets the delay between the soft and hard
channels, and this naturally leads to the observed plateau in the time lags at low
Fourier frequencies.

In future work, we plan to develop a more general Green's function in which the
injection occurs on a ring or a point, rather than on a spherical shell as in the
model considered here. As in the present paper, the resulting Fourier transform
of the time-dependent Green's function in the general case will allow us to investigate
a variety of seed photon energy distributions (e.g., blackbody or bremsstrahlung).
The additional geometric flexibility in the general model should allow us to further
improve the agreement between the model predictions and the data, hence providing
new insights into the structure of the scattering corona and the underlying
accretion disk. We also plan to examine scenarios in which the electrons
cool during the transient in response to the upscattering of the injected photons.
This may help to explain the soft time lags observed in some accreting black-hole
sources (e.g., Fabian et al. 2009).

The authors are grateful to the anonymous referee who provided a variety
of insightful comments that helped to strengthen and clarify the results presented here.

\section{Appendix~A}

In order to use the series expansions developed in Sections~3 and 4 to
represent the Green's functions for the time-averaged (quiescent) spectrum and for the
Fourier transform of the time-dependent spectrum, it is necessary to
establish the orthogonality of the various spatial eigenfunctions. In
this section, we present a global proof of orthogonality of the spatial
eigenfunctions for both the homogeneous case (utilizing the mirror inner
boundary condition) and for the inhomogeneous case (utilizing the dual
free-streaming boundary condition). First we define the generic spatial
ODE, encompassing Equations~(\ref{eq44}), (\ref{eq67}),
(\ref{eq83}), and (\ref{eq102}), by writing
\begin{equation}
{1 \over z^{2-\alpha}} \dfrac{d}{dz}\left(z^{2+\alpha}\dfrac{d\Gamma_n}
{dz}\right) + \eta^2\xi_n \Gamma_n(z) = 0
\ ,
\label{eq133}
\end{equation}
such that,
\begin{equation}
\alpha =
\begin{cases}
0, & \textrm{homogeneous (quiescent \& Fourier transform) ,} \\
1, & \textrm{inhomogeneous (quiescent \& Fourier transform) ,}
\end{cases}
\label{eq134}
\end{equation}
\begin{equation}
\Gamma_n(z) =
\begin{cases}
Y_n(z), & \textrm{homogeneous (quiescent \& Fourier transform) ,} \\
y_n(z), & \textrm{inhomogeneous (quiescent) ,} \\
g_n(z), & \textrm{inhomogeneous (Fourier transform) ,}
\label{eq135}
\end{cases}
\end{equation}
and
\begin{equation}
\xi_n =
\begin{cases}
\lambda_n, & \textrm{homogeneous (quiescent \& Fourier transform) ,} \\
\lambda_n, & \textrm{inhomogeneous (quiescent) ,} \\
\lambda_n+3i\tilde{\omega}z, & \textrm{inhomogeneous (Fourier transform) .}
\label{eq136}
\end{cases}
\end{equation}

To establish orthogonality, we multiply Equation~(\ref{eq133}) by
$\Gamma_m(z)$ and then duplicate it with the indices exchanged,
after which we subtract the second equation from the first, yielding
\begin{equation}
\Gamma_m\dfrac{d}{dz}\left(z^{2+\alpha}\dfrac{d\Gamma_n}{dz}\right)
- \Gamma_n\dfrac{d}{dz}\left(z^{2+\alpha}\dfrac{d\Gamma_m}{dz}\right)
= - \eta^2 z^{2-\alpha}(\xi_n-\xi_m) \Gamma_n(z)\Gamma_m(z)
\ .
\label{eq137}
\end{equation}
Next, we integrate by parts with respect to $z$ over the computational domain
$z_{\rm in} \le z \le 1$ to obtain, after simplification,
\begin{equation}
\left(z^{2+\alpha}\Gamma_m\dfrac{d\Gamma_n}{dz}-z^{2+\alpha}
\Gamma_n\dfrac{d\Gamma_m}{dz}\right)\bigg|_{z_{\rm in}}^{1}
= -\eta^2(\xi_n-\xi_m) \int_{z_{\rm in}}^{1}z^{2-\alpha}
\Gamma_n(z) \Gamma_m(z)\, dz
\ .
\label{eq138}
\end{equation}
The left-hand side of Equation~(\ref{eq138}) needs to be evaluated
separately for the homogeneous and inhomogeneous cases, since the
spatial boundary conditions are different in the two situations.
We consider each of these cases in turn below.

For the homogeneous cloud configuration, with $\alpha=0$ and
$z_{\rm in}=0$, the inner and outer spatial boundary conditions
can be written as (cf. Equations~(\ref{eq49}) and (\ref{eq51}))
\begin{equation}
\lim_{z \rightarrow 0} \ z^2 \, {d\Gamma_n \over dz} = 0 \ ,
\ \ \ \ \ \ \
\lim_{z \rightarrow 1} \left[{1 \over 3\eta} {d\Gamma_n \over dz}
+ \Gamma_n\right] = 0
\ .
\label{eq139}
\end{equation}
Likewise, in the inhomogeneous case, with $\alpha=1$, we can express
the inner and outer boundary conditions as (cf. Equations~(\ref{eq88})
and (\ref{eq90}))
\begin{equation}
\lim_{z \rightarrow z_{\rm in}} \left[\dfrac{z}{3\eta}\dfrac{d\Gamma_n}{dz}
- \Gamma_n\right] = 0
\ , \ \ \ \ \ \
\lim_{z \rightarrow 1} \left[\dfrac{z}{3\eta}\dfrac{d\Gamma_n}{dz}
+ \Gamma_n\right] = 0
\ .
\label{eq140}
\end{equation}
Using either the homogeneous or inhomogeneous boundary conditions given
by Equations~(\ref{eq139}) and (\ref{eq140}), respectively, we
find that the left-hand side of Equation~(\ref{eq138}) vanishes,
which establishes the required orthogonality of the
spatial eigenfunctions. The orthogonality condition can be written in
general as
\begin{equation}
\int_{z_{\rm in}}^{1}z^{2-\alpha}\,\Gamma_n(z)\,
\Gamma_m(z) \, dz = 0 \ , \ \ \ \ \ n \ne m
\label{eq141}
\ .
\end{equation}

\section{Appendix~B}

As shown in Section~5, the particular solution for the Fourier transform in
the case of bremsstrahlung, $F_{\rm brem}$, injection is given by
the convolution (see Equation (\ref{eq123}))
\begin{equation}
F_{\rm brem}(x,z,z_0,\tilde\omega) = \int_{x_{\rm abs}}^\infty
\Green(x,x_0,z,z_0,\tilde\omega)
\, A_0 \, x_0^{-1} e^{-x_0} N_0^{-1} \, dx_0
\ ,
\label{eq142}
\end{equation}
where $x_{\rm abs}$ is the dimensionless self-absorption cutoff energy, the
constant $A_0$ is given by Equation~(\ref{eq122b}), and the Fourier transform
Green's function, $\Green$, is given by Equations~(\ref{eq79}) and (\ref{eq119})
in the homogeneous and inhomogeneous cases, respectively. In general, we
can write $\Green$ in the generic form
\begin{equation}
\Green(x,x_0,z,z_0,\tilde\omega) = N_0 \, e^{(x_0-x)/2} (x x_0)^{-2}
\sum_{n=0}^{\infty}
M_{2,\lambda}(x_{\rm min})W_{2,\lambda}(x_{\rm max}) \, \mathscr{A}_n(z,z_0,\tilde\omega)
\ ,
\label{eq143}
\end{equation}
where $\xmin={\rm min}(x,x_0)$, $\xmax={\rm max}(x,x_0)$, and
$\mathscr{A}_n$ is a composite function containing the expansion
coefficients and the spatial eigenfunctions, given by
\begin{equation}
\mathscr{A}_n(z,z_0,\tilde\omega) =
\dfrac{e^{i \tilde{\omega} p_0} \eta^3}{4 \pi R^3
\Theta^4 (m_e c^2)^{3}}
\begin{cases}
\dfrac{\Gamma(\mu-3/2) Y_n(z_0) Y_n(z)}{\Gamma(1+2\mu)
\mathscr{I}_n} \, , & {\rm homogeneous} \, , \\ 
\dfrac{\Gamma(\sigma-3/2)g_n(z_0) g_n(z)}{\Gamma(1+2\sigma)
\, \eta^3 \mathscr{K}_n} \, , & {\rm inhomogeneous} \, .
\label{eq144}
\end{cases}
\end{equation}
In the homogeneous case, $\mu$ is computed using Equation~(\ref{eq72}), and
in the inhomogeneous case, $\sigma$ is computed
using Equation~(\ref{eq105}). Combining Equations~(\ref{eq142}) and (\ref{eq143}),
and reversing the order of summation and integration, we obtain
\begin{equation}
F_{\rm brem}(x,z,z_0,\tilde\omega) = A_0 \, e^{-x/2} x^{-2}
\sum_{n=0}^{\infty} \mathscr{A}_n(z,z_0,\tilde\omega)
B(\lambda,x)
\ ,
\label{eq145}
\end{equation}
where
\begin{equation}
B(\lambda,x) \equiv \int_{x_{\rm abs}}^{\infty} e^{-x_0/2}
x_0^{-3} M_{2,\lambda}(x_{\rm min})W_{2,\lambda}(x_{\rm max})
\, dx_0
\ ,
\label{eq146}
\end{equation}
and we set $\lambda=\mu$ to treat the homogeneous case, and we set
$\lambda=\sigma$ to treat the inhomogeneous case.

Our remaining task is to evaluate the integral function $B$
analytically, if possible. The expression for $B$ can be broken into
two integrals by writing, for $x \ge x_{\rm abs}$,
\begin{equation}
B(\lambda,x) = I_M(\lambda,x_0) \Bigg|_{x_{\rm abs}}^{x} W_{2,\lambda}(x)
+ I_W(\lambda,x_0) \Bigg|_{x}^{\infty} M_{2,\lambda}(x)
\ ,
\label{eq147}
\end{equation}
and, for $x \le x_{\rm abs}$,
\begin{equation}
B(\lambda,x) = I_W(\lambda,x_0) \Bigg|_{x_{\rm abs}}^{\infty} M_{2,\lambda}(x)
\ ,
\label{eq147b}
\end{equation}
where we have defined the indefinite integrals $I_M(\lambda,x_0)$ and
$I_W(\lambda,x_0)$ using
\begin{equation}
I_M(\lambda,x_0) \equiv \int e^{-x_0/2} x_0^{-3} M_{2,\lambda}(x_0) dx_0 \, , \ \ \ \
I_W(\lambda,x_0) \equiv \int e^{-x_0/2} x_0^{-3} W_{2,\lambda}(x_0) dx_0
\ .
\label{eq148}
\end{equation}
It is convenient to rewrite the Whittaker functions in the integrands for $I_M$ and $I_W$
using the Kummer function identities (Abramowitz \& Stegun 1970),
\begin{equation}
M_{\alpha,\beta}(z) = e^{-z/2}z^{\frac{1}{2}+\beta}M\Big(\frac{1}{2}+\beta-\alpha,
1+2\beta,z\Big)
\ ,
\label{eq150}
\end{equation}
\begin{equation}
W_{\alpha,\beta}(z) = e^{-z/2}z^{\frac{1}{2}+\beta}U\Big(\frac{1}{2}+\beta-\alpha,
1+2\beta,z\Big)
\ ,
\label{eq151}
\end{equation}
which yield
\begin{equation}
I_M(\lambda,x) = \int e^{-x} x^{b-a-5} M(a,b,x) dx \ , \ \ \ \
I_W(\lambda,x) = \int e^{-x} x^{b-a-5} U(a,b,x)dx 
\ ,
\label{eq152}
\end{equation}
where
\begin{equation}
a = \lambda - \frac{3}{2} \ , \ \ \ \ \ \ b=2 \, \lambda + 1
\ .
\label{eq153}
\end{equation}

The integral $I_W(\lambda,x)$ can be carried out analytically using Slater's
(1960) identity,
\begin{equation}
\int e^{-x} x^{b-a-2} U(a,b,x)dx = -e^{-x}x^{b-a-1}U(a+1,b,x)
\ .
\label{eq154}
\end{equation}
Integrating Equation~(\ref{eq152}) by parts once yields
\begin{equation}
\int x^{-3} e^{-x} x^{b-a-2} U(a,b,x)dx = -x^{-3} e^{-x} x^{b-a-1}
U(a+1,b,x)-3\int x^{-4}e^{-x} x^{b-a-1}U(a+1,b,x)dx
\ .
\label{eq155}
\end{equation}
Integrating by parts again gives
\begin{multline}
\int x^{-3} e^{-x} x^{b-a-2} U(a,b,x)dx = -x^{-3} e^{-x} x^{b-a-1}
U(a+1,b,x)-3\Big[-x^{-2}e^{-x} x^{b-a'-1}U(a'+1,b,x)\\
-2\int x^{-3}e^{-x}x^{b-a'-1}U(a'+1,b,x)dx\Big]
\ ,
\label{eq156}
\end{multline}
where $a'=a+1$. Integrating by parts a third time yields
\begin{multline}
\int x^{-3} e^{-x} x^{b-a-2} U(a,b,x)dx = -x^{-3} e^{-x} x^{b-a-1}
U(a+1,b,x)-3\Big\{-x^{-2}e^{-x} x^{b-a'-1}U(a'+1,b,x)\\
-2\Big[-x^{-1}e^{-x}x^{b-a''-1}U(a''+1,b,x)-\int x^{-2}e^{-x}
x^{b-a''-1}U(a''+1,b,x)dx\Big]\Big\}
\ ,
\label{eq157}
\end{multline}
where $a''=a'+1$. The remaining integral can be evaluated directly using
Equation~(\ref{eq154}) to obtain, after some algebra,
\begin{multline}
\int x^{-3} e^{-x} x^{b-a-2} U(a,b,x)dx = e^{-x}x^{b-a-4}
\Big[-U(a+1,b,x)+3U(a+2,b,x)\\
-6U(a+3,b,x)+6U(a+4,b,x)\Big]
\ .
\label{eq158}
\end{multline}
By converting the Kummer functions to Whittaker functions, we obtain
the final expression
\begin{equation}
I_W(\lambda,x) = e^{-x/2}x^{-2}\Big[- W_{1,\lambda}(x) + 3 W_{0,\lambda}(x)
- 6W_{-1,\lambda}(x) + 6W_{-2,\lambda}(x)\Big]
\ .
\label{eq159}
\end{equation}

Likewise, the integral $I_M(\lambda,x)$ in Equation~(\ref{eq152}) can be evaluated
using Slater's (1960) identity
\begin{equation}
\int e^{-x} x^{b-a-2} M(a,b,x) dx = \dfrac{e^{-x}x^{b-a-1}}{b-a-1}M(a+1,b,x)
\ .
\label{eq160}
\end{equation}
Following the same iterative procedure used to evaluate $I_W(\lambda,x)$, we
eventually arrive at the result
\begin{multline}
I_M(\lambda,x)= \dfrac{e^{-x} x^{b-a-4}}{b-a-1}\Bigg( M(a+1,b,x) + \dfrac{3}
{b-a-2}\bigg\{M(a+2,b,x) \\
+ \dfrac{2}{b-a-3}\Big[M(a+3,b,x)+\dfrac{1}{b-a-4} M(a+4,b,x)\Big]
\bigg\}\Bigg)
\ ,
\label{eq161}
\end{multline}
which can be rewritten in terms of the Whittaker functions as
\begin{equation}
I_M(\lambda,x) = \dfrac{x^{-2} e^{-x/2}}{\lambda+\frac{3}{2}} \Bigg(M_{1,\lambda}(x)
+ \dfrac{3}{\lambda+\frac{1}{2}} \bigg\{M_{0,\lambda}(x) + \dfrac{2}{\lambda-\frac{1}{2}}
\Big[M_{-1,\lambda}(x) + \dfrac{1}{\lambda-\frac{3}{2}} M_{-2,\lambda}(x)\Big] \bigg\}
\Bigg)
\ .
\label{eq162}
\end{equation}

Our final expression for the integral function $B(\lambda,x)$ is obtained by rewriting
Equations~(\ref{eq147}) and (\ref{eq147b}) as
\begin{equation}
B(\lambda,x) =
\begin{cases}
W_{2,\lambda}(x) [I_M(\lambda,x) - I_M(\lambda,x_{\rm abs})] - M_{2,\lambda}(x)
I_W(\lambda,x) \, , & x \ge x_{\rm abs} \, , \\
- M_{2,\lambda}(x) I_W(\lambda,x_{\rm abs}) \, , & x \le x_{\rm abs} \, ,
\end{cases}
\label{eq163}
\end{equation}
where $I_W(\lambda,x)$ and $I_M(\lambda,x)$ are evaluated using Equations~(\ref{eq159})
and (\ref{eq162}), respectively. We can now combine Equations~(\ref{eq144}) and
(\ref{eq145}) to express the bremsstrahlung injection Fourier transform $F_{\rm brem}$ as
\begin{equation}
F_{\rm brem}(x,z,z_0,\tilde\omega) =
\dfrac{e^{i \tilde{\omega} p_0} \eta^3 A_0 \, e^{-x/2}}{4 \pi R^3
\Theta^4 (m_e c^2)^{3} x^2}\sum_{n=0}^\infty
\begin{cases}
\dfrac{\Gamma(\mu-3/2) Y_n(z_0) Y_n(z)}{\Gamma(1+2\mu)
\mathscr{I}_n} \, B(\mu,x) \, , & {\rm homogeneous} \, , \\ 
\dfrac{\Gamma(\sigma-3/2)g_n(z_0) g_n(z)}{\Gamma(1+2\sigma)
\, \eta^3 \mathscr{K}_n} \, B(\sigma,x) \, , & {\rm inhomogeneous} \, ,
\label{eq164}
\end{cases}
\end{equation}
where $B(\mu,x)$ and $B(\sigma,x)$ are evaluated using Equation~(\ref{eq163}).

{}

\label{lastpage}

\end{document}